\documentclass[aps,jcp,twocolumn,groupedaddress]{revtex4}
\usepackage{graphicx,color,subfigure,amsmath}
\begin{document}

\title{Effect of chain stiffness on the dynamics and microstructure of crystallizable bead-spring polymer melts}
\author{Hong T. Nguyen}
\author{Robert S. Hoy}
\email{rshoy@usf.edu}
\affiliation{Department of Physics, University of South Florida, Tampa, FL, 33620}
\date{\today}

\begin{abstract}
We constrast the dynamics in model unentangled polymer melts of chains of three different stiffnesses:\ flexible, intermediate, and rodlike.  Flexible and rodlike chains, which readily solidify into close-packed crystals (respectively with randomly oriented and nematically aligned chains), display simple melt dynamics with Arrhenius temperature dependence and a discontinuous change upon solidification. 
Intermediate-stiffness chains, however, are fragile glass-formers displaying Vogel-Fulcher dynamical arrest, despite the fact that they also possess a nematic-close-packed crystalline ground state. 
To connect this difference in dynamics to the differing microstructure of the melts, we examine how various measures of structure, including cluster-level metrics recently introduced in studies of colloidal systems, vary with chain stiffness and temperature.
No clear static-structural cause of the dynamical arrest is found.
However, we find that the intermediate-stiffness chains display qualitatively different dynamical heterogeneity.
Specifically, their stringlike correlated motion (cooperative rearrangement) is correlated along chain backbones in a way not found for either flexible or rodlike chains.
This activated ``crawling'' motion seems to be the cause of the dynamical arrest observed for these systems, and shows that the factors controlling the crystallization vs.\ glass formation competition in polymers likely depend nonmonotonically on chain stiffness.
\end{abstract}
\maketitle

\section{Introduction}
\label{sec:intro}

The competition between crystallization and glass-formation is of longstanding scientific interest.\cite{frank52,royall15}
It naturally depends in detail on local fluctuations away from equilibrium liquid structure.  
Since these fluctuations are both small (in lengthscale) and short-lived except at temperatures very near solidification, they are difficult to characterize in experiments.
Moreover, unlike crystal nucleation, vitrification has historically been difficult to associate with local structural change.
However, recent computer simulations that have focused on understanding solidification at a microscopic level\cite{royall15,shintani06,tanaka10,russo12,taffs13,malins13a,malins13b} have led to much progress in our understanding of the crystallization-glass formation (CF-GF) competition.
These studies have identified key structural features within liquids, at the level of \textit{clusters}  of $\sim10$ particles, that dramatically influence solidification.
Differently structured clusters have different thermodynamic stabilities, and thus both their formation propensities and characteristic lifetimes vary differently with temperature.
Recent analyses of such differences have identified long-lived, stable amorphous clusters that strongly promote glass-formation; clusters that are fivefold symmetric\cite{shintani06} and/or are subsets of icosahedra\cite{malins13a,malins13b} are particularly effective glass-promoters.
Other work\cite{shintani06,kawasaki07,tanaka10,leocmach12,russo12} has found that clusters possessing bond-orientational order compatible with the stable crystal can be present and long-lived above the solidification temperature, and that these dynamically slow regions of ``medium range crystalline order'' (MRCO\cite{shintani06}) promote crystallization.

To date, most analyses of this type have been carried out only for relatively simple models of atomic or colloidal systems, e.g.\ the Wahnstr{\"o}m\cite{wahnstrom91} and Kob-Andersen\cite{kob94} binary Lennard-Jones mixtures.
In particular, they have not been extended to polymers.\cite{collpol}
Polymers exhibit a particularly complicated solidification behavior that is strongly thermal-history-dependent\cite{keller68} and also depends in considerable detail on their rich liquid-state \textit{dynamics}.
For example, the locally nematic alignment of chain molecules which typically precedes crystallization is rate-limited by the slow dynamics of entangled melts.\cite{doi86,strobl09}.
If this alignment cannot occur because the cooling rate is too high, the system will glass-form.
Both the alignment propensity and single-chain dynamics in turn depend strongly on the stiffness of the polymer chains.\cite{faller99,faller01} 
Understanding the CF-GF competition therefore requires understanding both the structure and dynamics of the parent liquid.

The goal of this paper is to improve our understanding of how the microstructure and dynamics of polymer melts depend on temperature and chain stiffness, using molecular simulations and with an eye towards improving our understanding of the solidification process.
For simplicity, we employ a recently developed coarse-grained model that has been used to study the CF-GF competition,\cite{hoy13,nguyen15} and unentangled chains.
By tuning a single model parameter (the chain stiffness $k_b$), we show that marked differences in solidification behavior coincide with marked differences in melt dynamics.
Flexible chains that form random-walk-close-packed (RWCP) crystals upon solidification,\cite{nguyen15} wherein monomers are close-packed but chains adopt random-walk-like configurations, exhibit a simple melt dynamics that remains ``fast'' down to solidification.
Stiff chains that form nematic melts and solidify into nematic close-packed (NCP) crystals,\cite{nguyen15} wherein monomers are close-packed while chains adopt rod-like configurations and are aligned nematically, also exhibit such ``simple, fast'' melt dynamics.
In sharp contrast, for an intermediate chain stiffness that produces glass-formation upon cooling,\cite{nguyen15} systems exhibit the dynamics of fragile glass-formers, including Vogel-Fulcher dynamical arrest, even though their ground state is NCP.
We attempt to connect these differences to static structure using various tools recently developed in studies of colloidal glass- and crystal-formers, that have, however, not yet been applied to polymeric systems.
In contrast to data for some colloidal systems,\cite{malins13b,royall15} cluster-level measures of static and dynamic structure do not exhibit a clear signature differentiating our crystallizing and glassforming polymer melts.
Instead, we find that the cause of the dynamical slowdown producing glass-formation in intermediate-stiffness systems is that the stringlike cooperative motion\cite{donati98,donati99} associated with dynamical heterogeneity becomes coordinated along chain backbones in a fashion not found for either flexible or stiff chains.

\section{Model and Methods}
\label{sec:methods}

\subsection{Molecular dynamics simulations}
\label{subsec:MDsims}

Our study employs the soft-pearl-necklace polymer model used in Refs.\ \cite{hoy13,nguyen15}.
It is based on the semiflexible version of the widely-used Kremer-Grest (KG) bead-spring model \cite{kremer90,auhl03}, but uses a different potential for covalent backbone bonds.
All monomers have mass $m$ and interact via the truncated and shifted Lennard-Jones potential
\begin{equation}
U_{LJ} = 4\epsilon\left[\left(\displaystyle\frac{\sigma}{r}\right)^{12} - \left(\displaystyle\frac{\sigma}{r}\right)^{6} - \left(\displaystyle\frac{\sigma}{r_c}\right)^{12} + \left(\displaystyle\frac{\sigma}{r_c}\right)^{6}\right],
\label{eq:LJpot}
\end{equation}
where $\epsilon$ is the intermonomer binding energy and $r_c$ is the cutoff radius. 
Attractive Van der Waals interactions are included by setting $r_c = 2^{7/6}\sigma$.

Covalent bonds connecting consecutive monomers along chain backbones are modeled using the harmonic potential
\begin{equation}
U_c(\ell) = \displaystyle\frac{k_c}{2}\left(\ell-a\right)^2,
\label{eq:Ubond}
\end{equation}
where $\ell$ is bond length, $a$ is monomer diameter, and $k_c = 600\epsilon/a^2$ is the bond stiffness.
For this value of $k_c$, the energy barrier for chain crossing is at least $50k_BT$ over the whole temperature range considered herein.

Angular interactions between three consecutive beads along chain backbones are modeled by the standard potential \cite{auhl03}
\begin{equation}
U_b(\theta) = k_b(1 - cos(\theta)),
\label{eq:Ubend}
\end{equation}
where $\theta_i$ is the angle between consecutive bond vectors $\vec{b}_i$ and $\vec{b}_{i+1}$; here $\vec{b}_i = \vec{r}_{i+1}-\vec{r}_i$ and $\vec{r}_i$ is the position of bead $i$.
Note that $\theta$ is zero and $U_b$ is minimized for straight trimers.
In this paper we consider three representative chain stiffnesses identified in Ref.\ \cite{nguyen15}: flexible ($k_b = 0$), intermediate ($k_b = 4\epsilon$), and stiff ($k_b = 12.5\epsilon$).

The KG model is a good glass-former \cite{bennemann99,abrams01} largely because its equilibrium backbone bond length $\ell_0$ is incommensurable with its equilibrium nearest neighbor distance for nonbonded neighbors, $r_0$.
Specifically, it has $\ell_0 = 0.96a$ and $r_0 = 2^{1/6}a$. 
In contrast, the current model makes these lengths commensurable ($\ell_0 = r_0 = a$).
We obtain polymer chains with $\ell_0 = r_0 = a$ by setting $\sigma = 2^{-1/6}a$.
This property gives it a unique, well-defined ground state for $k_b > 0$: the nematic and close-packed (NCP) crystal.\cite{nguyen15} 

We study this model using molecular dynamics (MD) simulations of cooling from high to low $T$, as well as constant-temperature melt dynamics.
All simulations are performed using the LAMMPS\cite{plimpton95} MD package.
All systems are composed of $N_{ch} = 500$ chains, each with $N=25$ monomers. 
These chains are unentangled. 
Periodic boundaries are applied along all three directions of cubic simulation cells. 
Initial states are well-equilibrated melts at temperatures well above their ($k_b$-dependent) solidification temperatures.
Temperature and pressure are controlled using a Nos{\'e}-Hoover thermostat and barostat. 
After equilibration at zero pressure, states are cooled (also at zero pressure) at rates $(k_B/u_0)|\dot{T}| = 10^{-6}/\tau_{LJ}$, $10^{-5}/\tau_{LJ}$, and $10^{-4}/\tau_{LJ}$, to $T=0$; here $\tau_{LJ}$ is the Lennard-Jones time unit $\sqrt{ma^2/\epsilon}$. 

Here we present results from these cooling runs, which are the same as those used in Ref.\ \cite{nguyen15}, and also from $NPT$ melt dynamics runs.
The latter are prepared by taking snapshots at different temperatures $T_i$ from the $|\dot{T}| = 10^{-6}/\tau$ cooling run, and allowing their structure to relax at zero pressure and $T=T_i$, producing equilibrium liquids and metastable supercooled liquids at the various $T_i$.
Then systems are integrated forward in time at fixed (zero) pressure and temperature for up to a few million $\tau_{LJ}$.
The damping times of the thermostat and barostat are $(\tau_{T},\tau_{P}) = (\tau_{LJ}, 10\tau_{LJ})$ during the cooling runs and $(\tau_{T},\tau_{P}) = (10\tau_{LJ}, 100\tau_{LJ})$ during the $NPT$ dynamics runs.,
In all runs, the MD timestep used is $\delta t = \tau_{LJ}/200$.
In the remainder of the paper, we present all quantities in Lennard-Jones units.

To characterize $T$- and $k_b$-dependent structure, we monitor the packing fraction $\phi(T)$, the fraction of atoms with close-packed order $f_{cp}(T)$, and the \textit{bond}-scale nematic order parameter\cite{luo11} $\mathcal{O}(T)$ during the cooling runs.
Here $\phi(T) = \pi\rho(T)/6$ where $\rho$ is monomer number density, $f_{cp}$ is determined by Characteristic Crystallographic Element (CCE) analysis \cite{cce09}, and $\mathcal{O}$ is given by\cite{luo11}
\begin{equation}
\mathcal{O}=\sqrt {\frac{3}{2} \rm{Tr}( q^2)},
~q_{\alpha\beta}= \biggl\langle \hat{b}_{\alpha}\hat{b}_{\beta} -\frac{1}{3} \delta_{\alpha\beta}  \biggl\rangle .
\label{eq:nemord}
\end{equation}
Here, Tr is the trace operator, $\left< \cdots \right>$ denotes averaging over all normalized bond vectors $\vec{b}$ in each sub-cell followed by averaging over all subcells (cubic cells of side length $2-3a$) in the simulation,\cite{luo11,nguyen15} and $\hat{b}_{\alpha}$ and $\hat{b}_{\beta}$ are Cartesian components of $\vec{b}/|\vec{b}|$.
$\mathcal{O} = 1$ corresponds to perfect alignment of bonds within subcells, while $\mathcal{O} = 0$ corresponds to random bond orientation.

To monitor melt dynamics, we calculate the self-intermediate scattering function $S(q_{peak},t)$ for the $NPT$ runs:
\begin{equation}
S(q_{peak},t) = \left<\frac{1}{N_{tot}N_{\vec{q}}} \sum_{j}^{N_{\vec{q}}} \sum_{i}^{N_{tot}}e^{-i\vec{q}_j\cdot(\vec{r}_i(t)-\vec{r}_i(0))}\right>,
\label{eq:soq}
\end{equation}
where $q_{peak}$ is obtained by fitting a Gaussian function to the first peak of the structure factor $S(q)$, and $N_{\vec{q}} \sim N_{tot}$ is the number of wavevectors with magnitude in the range $[q_{peak}-0.1/a,q_{peak}+0.1/a]$.
Following standard practice,\cite{sastry98} the alpha time $\tau_{\alpha}$ is first defined as the time at which $S(q_{peak},t)$ reaches $1/e$.

\subsection{TCC analyses}

The Topological Cluster Classification (TCC)\cite{malins13d} is a method for identifying inhomogeneous local structure in condensed matter.
It groups particles into $N$-body ``clusters'' and then distinguishes differently structured clusters by their differing interparticle topology.
Of particular interest are ``locally favored structures''\cite{royall08}, the ground-state $N$-particle clusters, but other structures of higher energy are often also present in large numbers in nonequilibrium systems as well as in systems at $T$ sufficiently high that thermal fluctuations are important.\cite{malins13a,malins13b,malins13d}
Here we employ TCC to track the formation propensities and lifetimes of various microstructural motifs within our systems, during both the cooling and $NPT$ dynamics runs.
The idea is to connect any differences in the dynamics to differences in microstructure.  

\begin{figure}[htbp]
\includegraphics[width=2.75in]{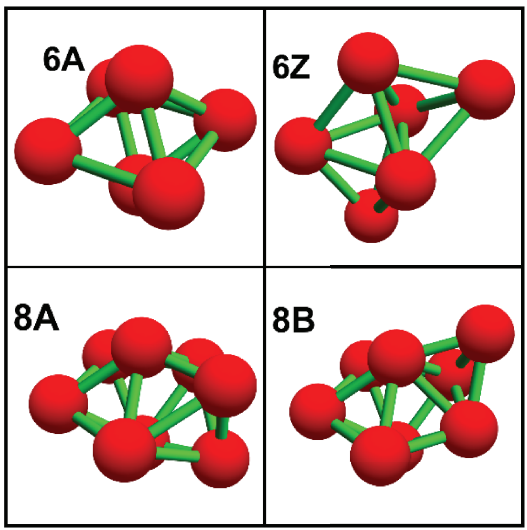}
\caption{The four clusters of primary interest here, as identified by TCC.  The cluster-identification notation follows Refs.\ \cite{doye95,malins13a,malins13b,malins13d}.}
\label{fig:fourclusts}
\end{figure}

During the cooling runs, we monitor the fractions $f_X(T)$ of particles belonging to at least one cluster of type $X$.
These show how microstructure varies with temperature; of particular interest are changes in the ratios $f_X/f_Y$ of differently structured clusters.
We use the same procedures detailed in Ref.\ \cite{malins13d}, and identify $f_X$ for many different clusters.
Figure \ref{fig:fourclusts} shows four clusters that are of particular interest here, denoted (according to the naming scheme of Doye et.\ al.\ \cite{doye95}) $6A,\ 6Z,\ 8A,\ \rm{and}\ 8B$.
$6A$ is the octahedron and is compatible with crystalline close-packing.
$6B$ is the other LFS for $N=6$, possesses a partial fivefold-symmetric structure, and is a subset of the icosohedron and therefore incompatible with crystallization.
The two $N=8$ clusters are similarly ``amorphous''; $8B$ is a subset of the icosohedron, while $8A$ is not.

During the dynamics runs, we monitor the lifetimes of the various clusters.
A cluster is considered ``alive'' at time $t$ if the same $N$ atoms formed a cluster at time zero.
We monitor
\begin{equation}
A_X(t) = \displaystyle\frac{1}{N_X(0)} \sum_{i=1}^{N_X(0)} G_{X,i}(0,t)
\label{eq:aoft}
\end{equation}
where $G_X(0,t)$ is unity if the same $N$ atoms make up an $X$-cluster at times $t' = 0$ and $t' = t$, and zero otherwise.  
$N_X(0)$ is the number of $X$-clusters at time $t'=0$, so $A(0) = 1$.
$A_X(t)$ decreases nearly monotonically (to zero at large $t$) since monomers diffuse away from each other in the melt, only rarely returning to their original positions.
The $T$-dependent cluster lifetimes $\tau_X$ are determined by identifying $A_X(2\tau_X) = 1/e^2$ \cite{why2tau}.

\section{Results}
\label{sec:results}

\subsection{Evolution of structure during cooling}

\begin{figure}[htbp]
\includegraphics[width=2.75in]{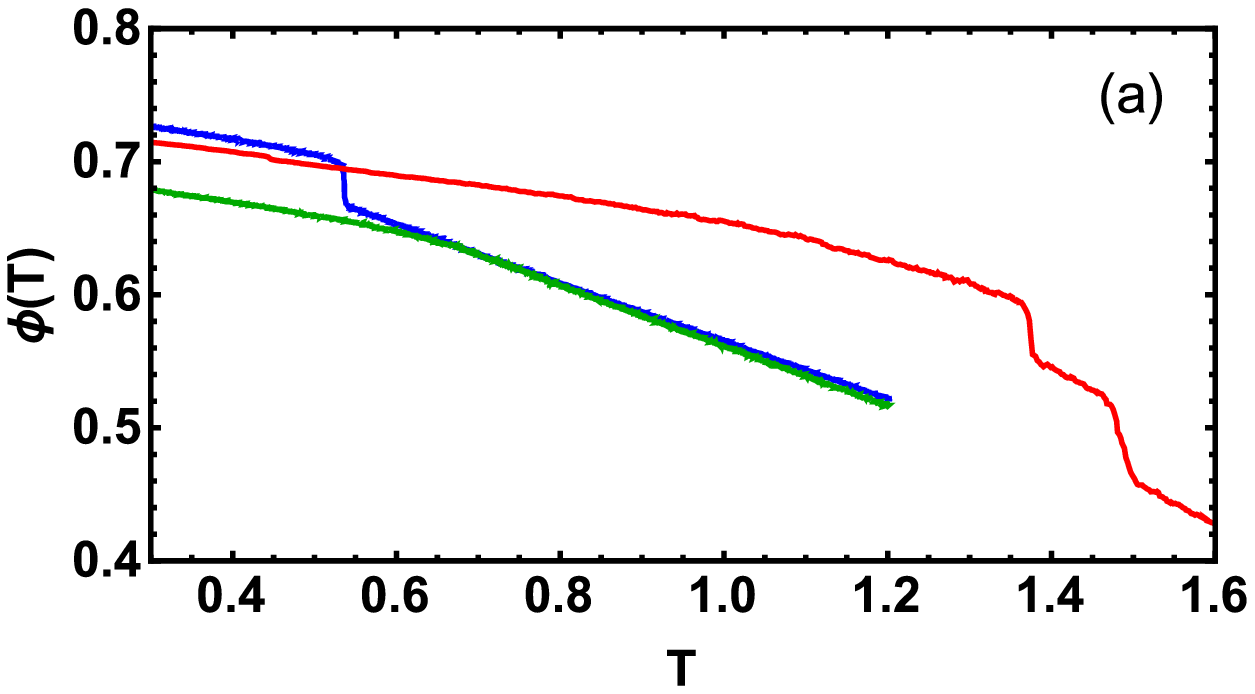}
\includegraphics[width=2.75in]{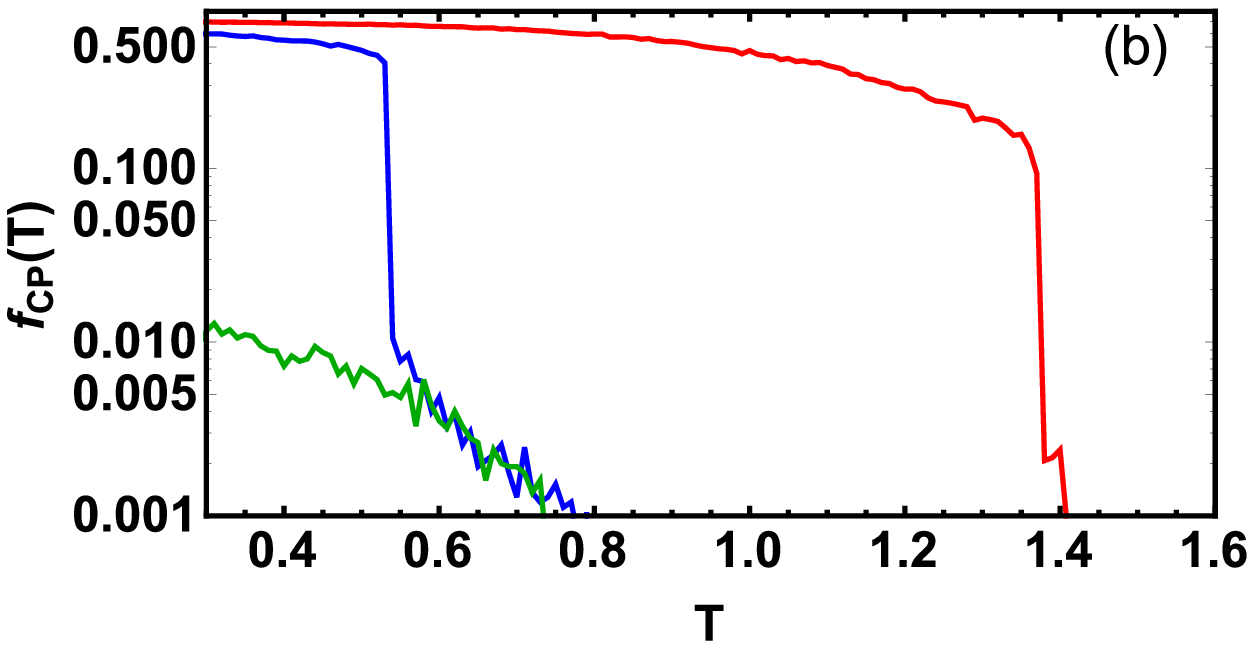}
\includegraphics[width=2.75in]{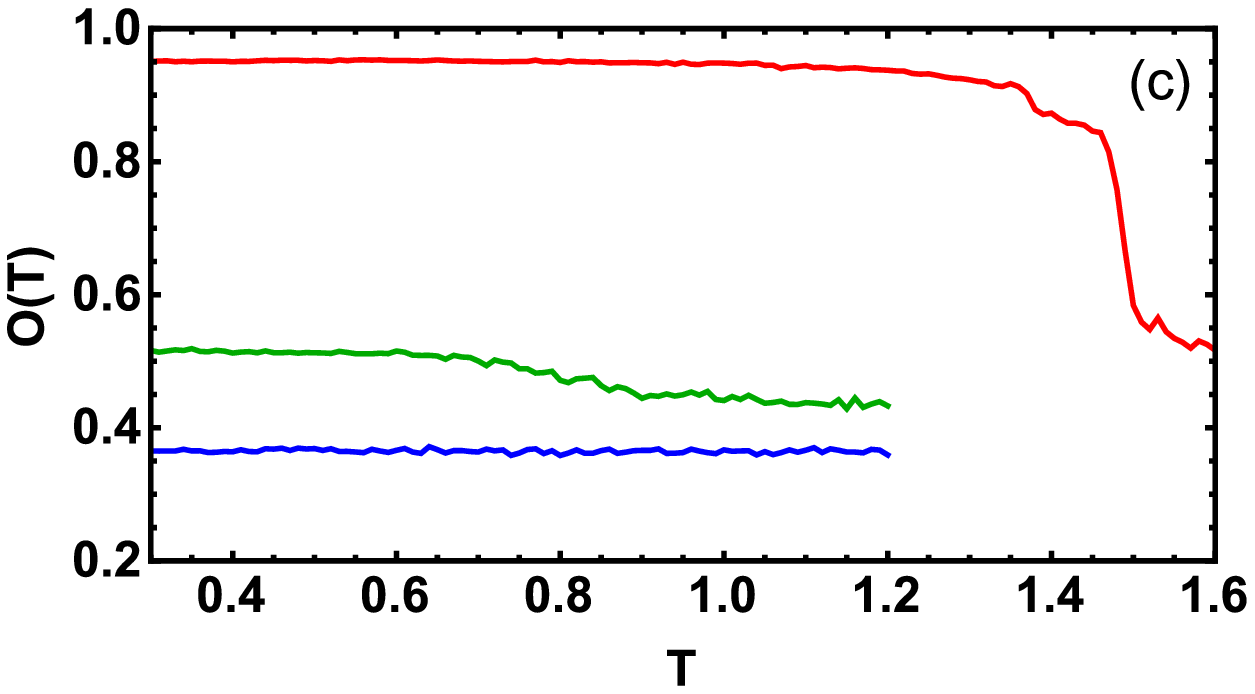}
\caption{Gross measures of structure across the liquid-solid transition for flexible ($k_{bend} = 0$; blue lines), intermediate-stiffness ($k_{bend}=4\epsilon$; green lines), and stiff ($k_{bend}=12.5\epsilon$; red lines) systems during $|\dot{T}|=10^{-6}/\tau_{LJ}$ quenches.  Panel (a): packing fraction $\phi(T)$.  Panel (b): fraction of close-packed sites\cite{cce09} $f_{cp}(T)$.  Panel (c): nematic order parameter $\mathcal{O}(T)$ (Eq.\ \ref{eq:nemord}).}
\label{fig:grossvsT}
\end{figure}

We now present basic results from the cooling runs, in order to place the dynamics results that follow in the context of the CF-GF competition for these systems.
Figure \ref{fig:grossvsT} shows the evolution of several measures of structure during $|\dot{T}|=10^{-6}$ cooling runs.
These results were also presented in Ref.\ \cite{nguyen15}, but are represented here to illustrate these systems' very different solidification behavior. 
Panel (a) shows the packing fraction $\phi(T)$.
At very high $T$, results for all systems fall on a universal curve corresponding to isotropic fluids.
Flexible ($k_b = 0$) and stiff ($k_b = 12.5\epsilon$) chains show sharp, first-order-like transitions upon crystallization, respectively at $T_s \simeq 0.56$ and $T_s \simeq 1.40$.\cite{footTs}
Stiff chains show another transition, from isotropic to nematic fluids, at $T_{ni} \simeq 1.52$; density increases as chains nematically align.
In contrast, intermediate-stiffness ($k_b = 4\epsilon$) chains show characteristically glassy behavior wherein only the slope $\partial \phi/\partial T$ changes noticeably upon solidification at $T_s \simeq 0.60$.

These differences are reinforced by examining the fraction  $f_{cp}(T)$ of monomers possessing locally close-packed environments.\cite{cce09}
For flexible and stiff chains, $f_{cp}$ increases sharply to a large value at $T_s$, as close-packed crystalline order develops.
In contrast, $f_{cp}$ for intermediate-stiffness chains increases only slightly as $T$ decreases and remains small even at $T=0$; this system forms an amorphous glassy state.\cite{nguyen15}

Another $k_b$-dependent difference is illustrated by examining the bond-scale nematic order $\mathcal{O}(T)$.
For flexible chains, nearby chain segments remain nearly randomly oriented; the finite value of $\mathcal{O}$ arises partially from the pearl-necklace structure \cite{faller99} and partially from the finite size of the subcells used to calculate $\mathcal{O}$ (Section \ref{subsec:MDsims}).
Stiff chains show two transitions: the isotropic-nematic transition at $T_{ni}$, and crystallization at $T_s$.
For temperatures slightly above solidification, the flexible and stiff melts possess very different structure; the former are isotropic while the latter are nematic.
Intermediate-stiffness chains show (as expected\cite{faller99}) intermediate behavior; some local nematic order is present at high $T$, and increases slightly upon cooling as chains uncoil and locally align.
However, this order is only short-ranged, in sharp contrast to stiff-chain systems where a single nematic domain spans the simulation cell.\cite{nguyen15}

\subsection{Self-intermediate scattering function}

\begin{figure}[htbp]
\includegraphics[width=2.75in]{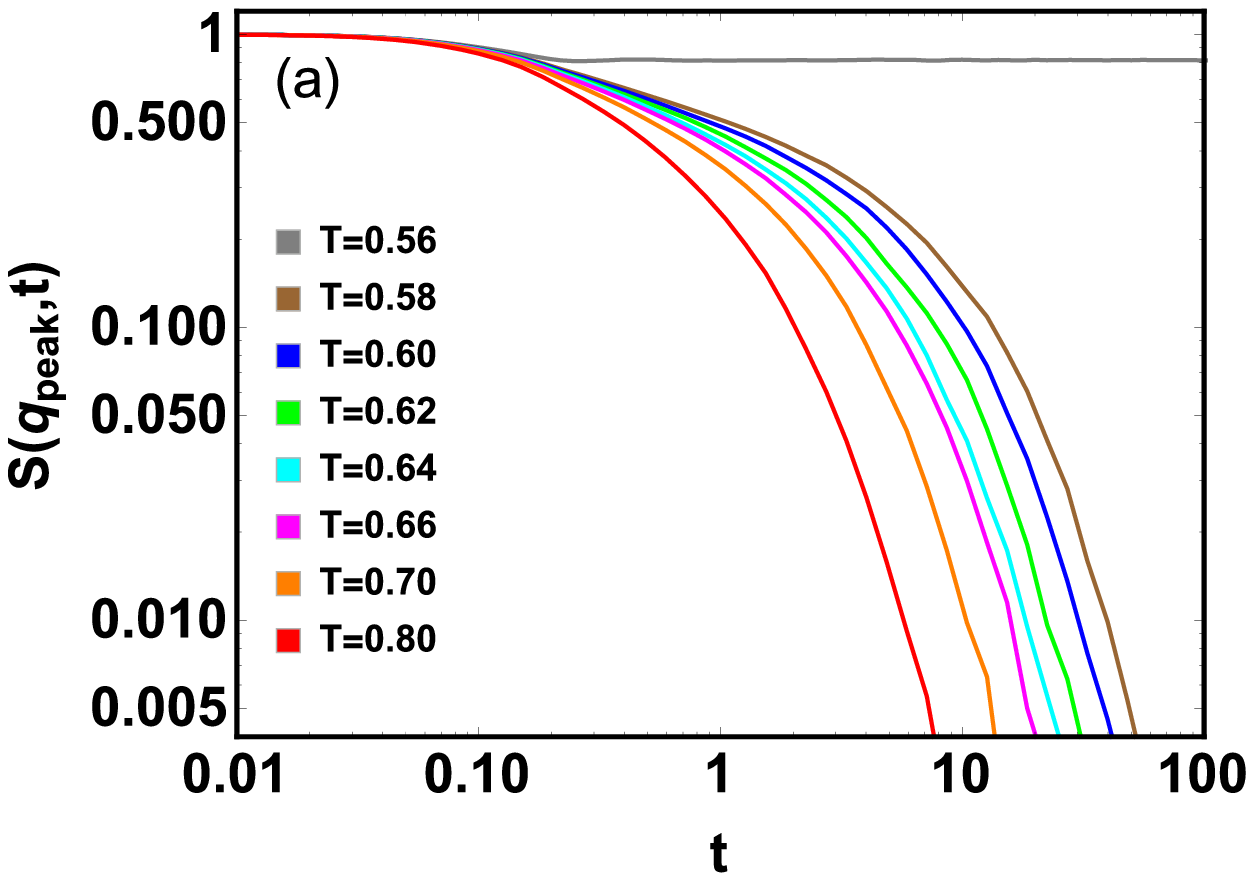}
\includegraphics[width=2.75in]{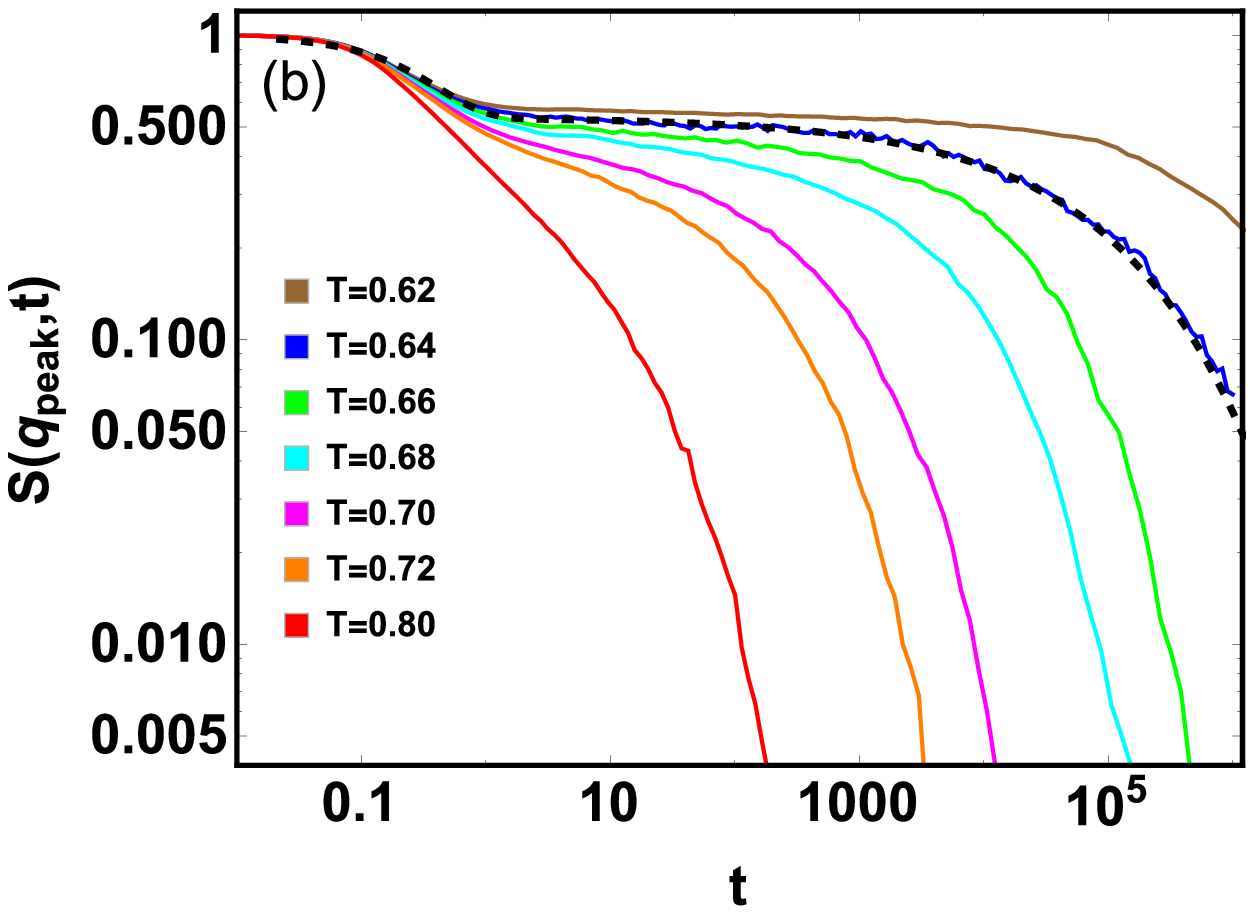}
\includegraphics[width=2.75in]{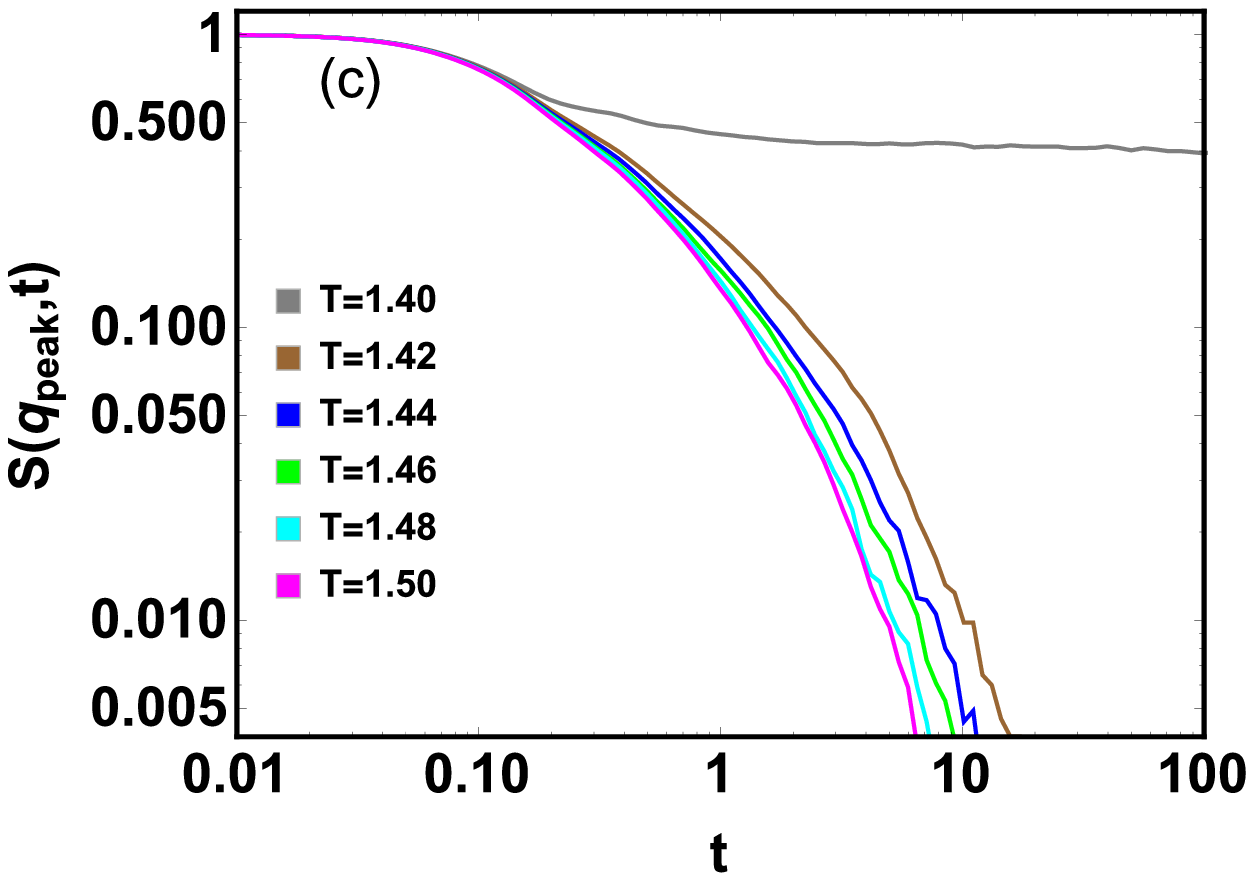}
\includegraphics[width=2.75in]{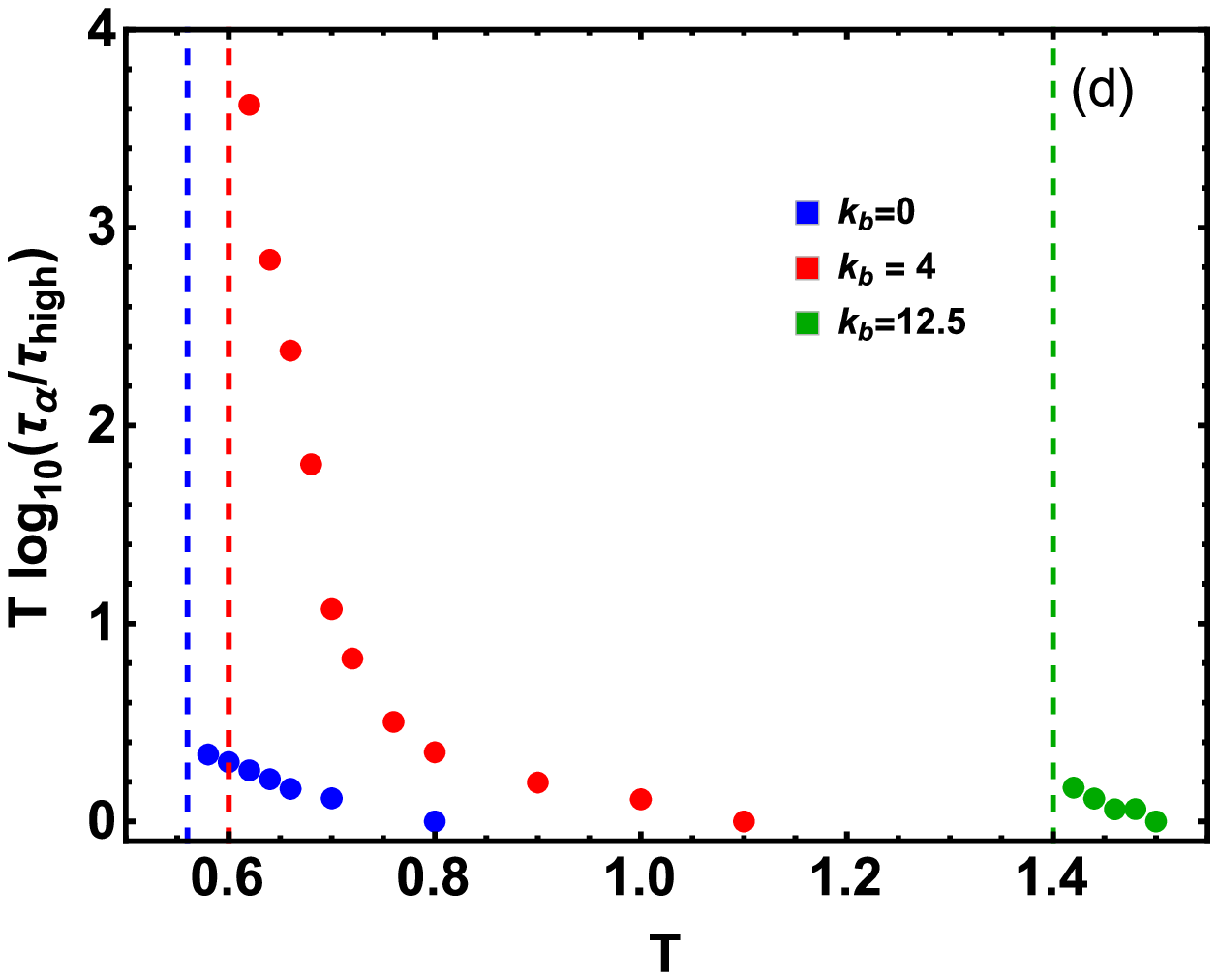}
\caption{Self-intermediate scattering function $S(q_{max},t)$ from NPT runs at various $T$ for (a) $k_{bend} = 0$,  (b) $k_{bend} = 4\epsilon$,   (c) $k_{bend} = 12.5\epsilon$. The black dashed line in (b) shows a fit to Eq.\ \ref{eq:alphabeta}.  Panel (d) shows results for $T \log_{10}(\tau_{\alpha}/\tau_{high})$; horizontal data at zero would indicate ideal Arrhenius behavior.  Vertical dashed lines in panel (d) show values of $T_s$.}
\label{fig:sqt}
\end{figure}

We now shift focus to comparing these systems' constant-temperature melt dynamics. 
Figure \ref{fig:sqt} shows the self-intermediate scattering function $S(q_{peak},t)$ for the three stiffnesses at various temperatures above solidification.
The relaxation of flexible- and stiff-chain melts (panels a, c) is ``fast and simple''; the decay of $S(q_{peak},t)$ is close to the single-exponential form typical for simple liquids.\cite{hansen86}
The slight deviations from single-exponential relaxation likely result either from the underlying Rouse dynamics of chains\cite{doi86,kremer90} or from $\alpha$ and $\beta$ relaxations occurring on timescales that are not well-separated.\cite{hansen86}
Relaxation in these systems is fast; $\tau_{\alpha}$ increases only to $\sim 100\tau_{LJ}$ for temperatures as low as $0.02$ above $T_s$. 
Panel (d) shows that its temperature dependence is almost Arrhenius; $T \log_{10}(\tau_{\alpha}/\tau_{high})$, where $\tau_{high}$ is equal to $\tau_{\alpha}$ at the highest tested $T$, remains less than $\sim 1/3$ over the studied temperature ranges.
Note that there is a sharp, discontinuous change in dynamics upon crystallization; gray curves in these panels correspond to systems that crystallized during sample preparation.

Panel (b) shows that intermediate-stiffness chains exhibit markedly different relaxation.
$S(q_{peak},t)$ is well-fit by the classical two-step stretched-exponential form
\begin{equation}
F(q,t)=(1-A) exp(-t/\tau_{\beta}) + A exp(-(t/\tau_{\alpha}^{F})^{\beta})
\label{eq:alphabeta}
\end{equation}
where $\tau_{\beta}$ and $\tau_{\alpha}^F$ are the slow and fast relaxation times, $A$ is the Debye-Waller factor, and $\beta < 1$ is the Kohlrausch-Williams-Watts (KWW) stretching parameter.   
We find that $\tau_{\alpha}^F$ increases by several orders of magnitude as $T-T_s$ decreases from $0.2$ to $0.02$, and its temperature dependence is well fit by the Vogel-Fulcher form
\begin{equation}
\tau_{\alpha}^F(T) = B \exp\left(\displaystyle\frac{DT_0}{T-T_0}\right)
\label{eq:VF}
\end{equation}
with $D \simeq 16.2$ and $T_0 \simeq 0.43$.
Panel (d) shows that non-Arrhenius relaxation sets in at $T \simeq 0.8$; below this $T$, $T \log_{10}(\tau_{\alpha}/\tau_{high})$ increases sharply as $T_s$ is approaches, with $\tau_{\alpha}$ increasing s by $\sim 3.5$ orders of magnitude more than would be expected for Arrhenius temperature dependence.
Such strongly non-Arrhenius behavior is typical of systems possessing dynamics strongly influenced by their energy landscapes.\cite{debenedetti01}
In short, intermediate-stiffness melts exhibit the dynamics of fragile glassformers.\cite{debenedetti01}

It is remarkable that intermediate-stiffness systems show this dramatic dynamical slowdown when both more-flexible and stiffer chains do not.
The observed slowdown suggests that equilibrium crystallization for $k_b = 4\epsilon$ systems would (hypothetically) occur at $T \simeq 0.8$.
Dynamical arrest is often found in systems with avoided crystallization transitions,\cite{royall15} and the behavior of our intermediate-stiffness system is consistent with that of a deeply supercooled liquid.

\begin{figure}[htbp]
\includegraphics[width=2.75in]{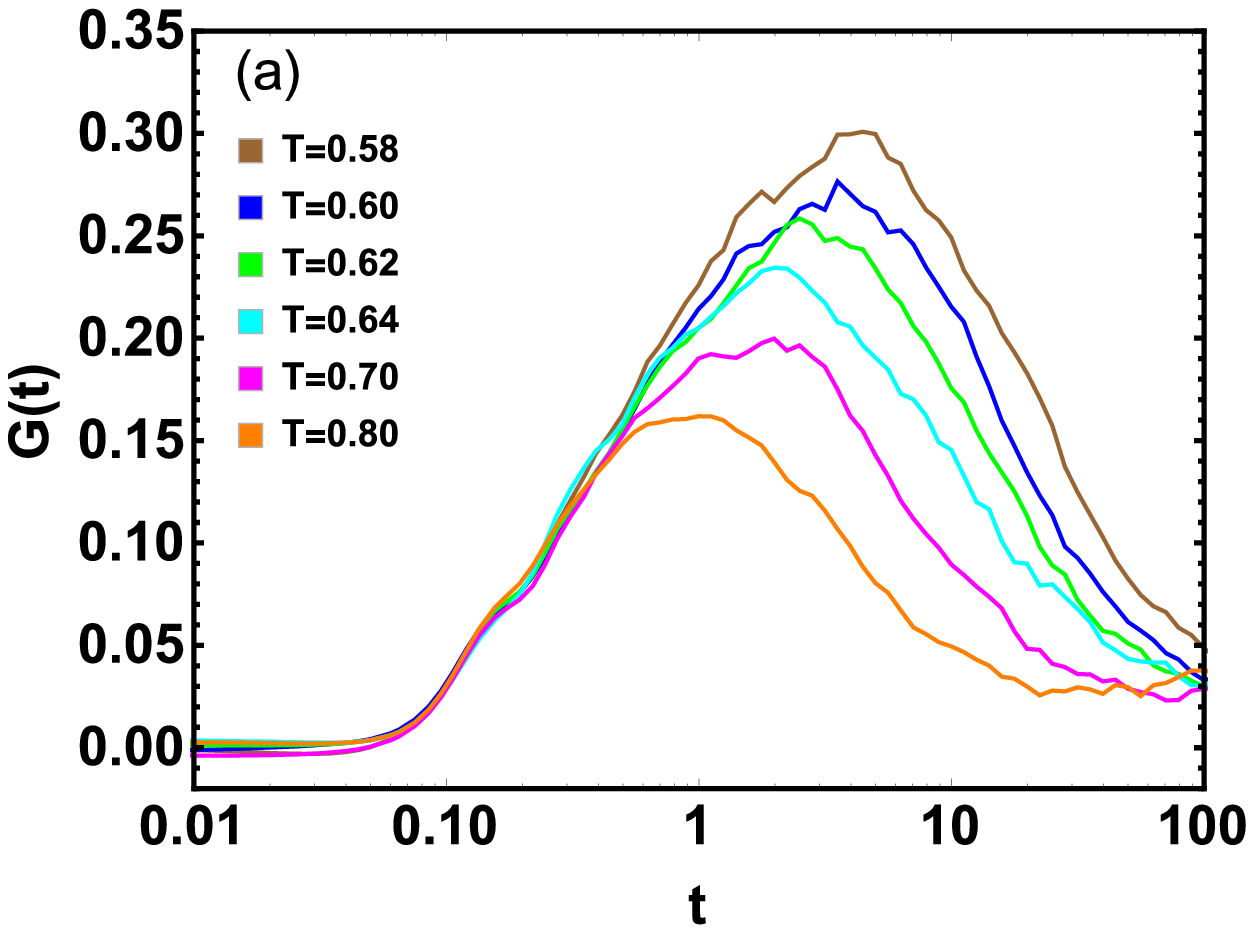}
\includegraphics[width=2.75in]{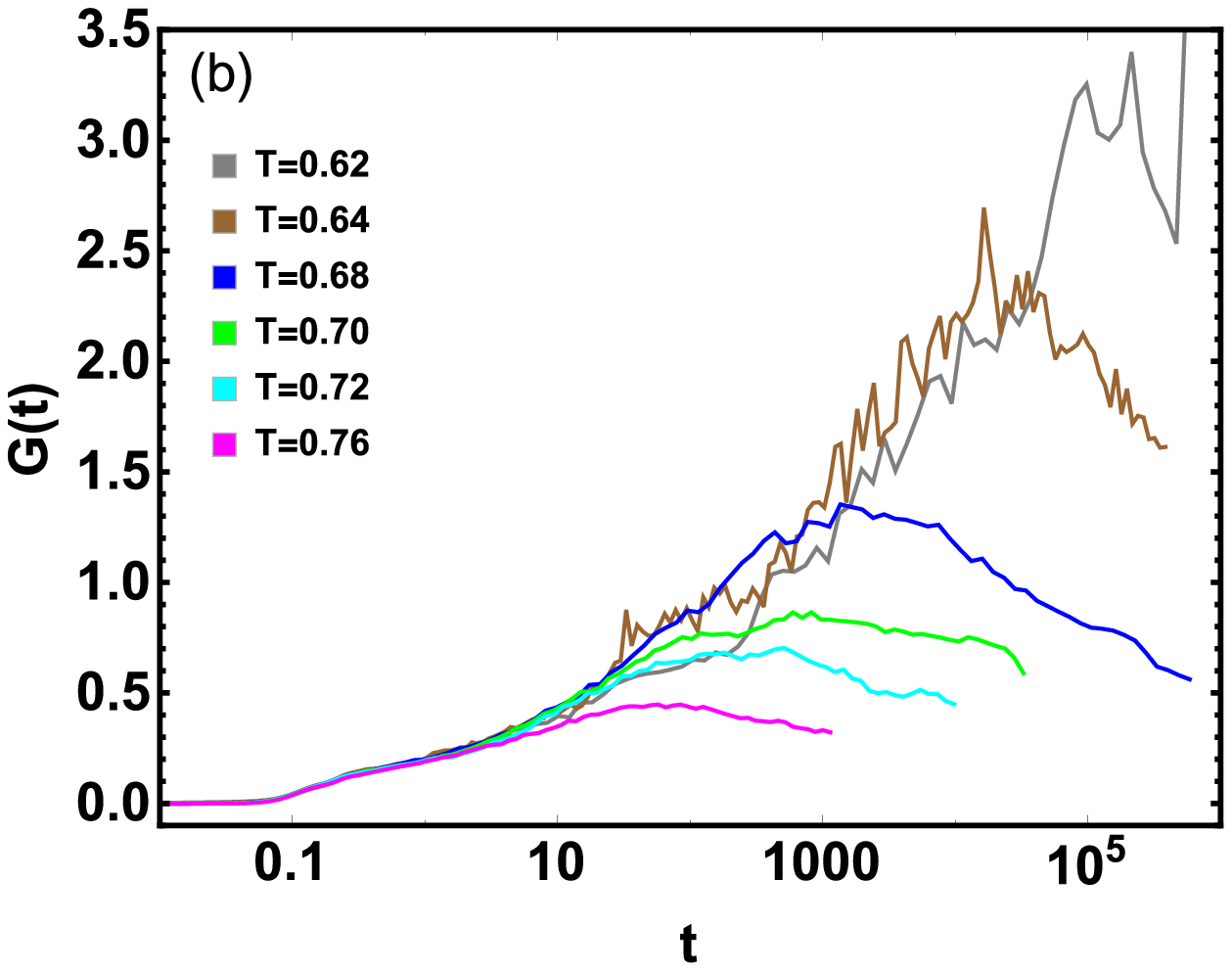}
\includegraphics[width=2.75in]{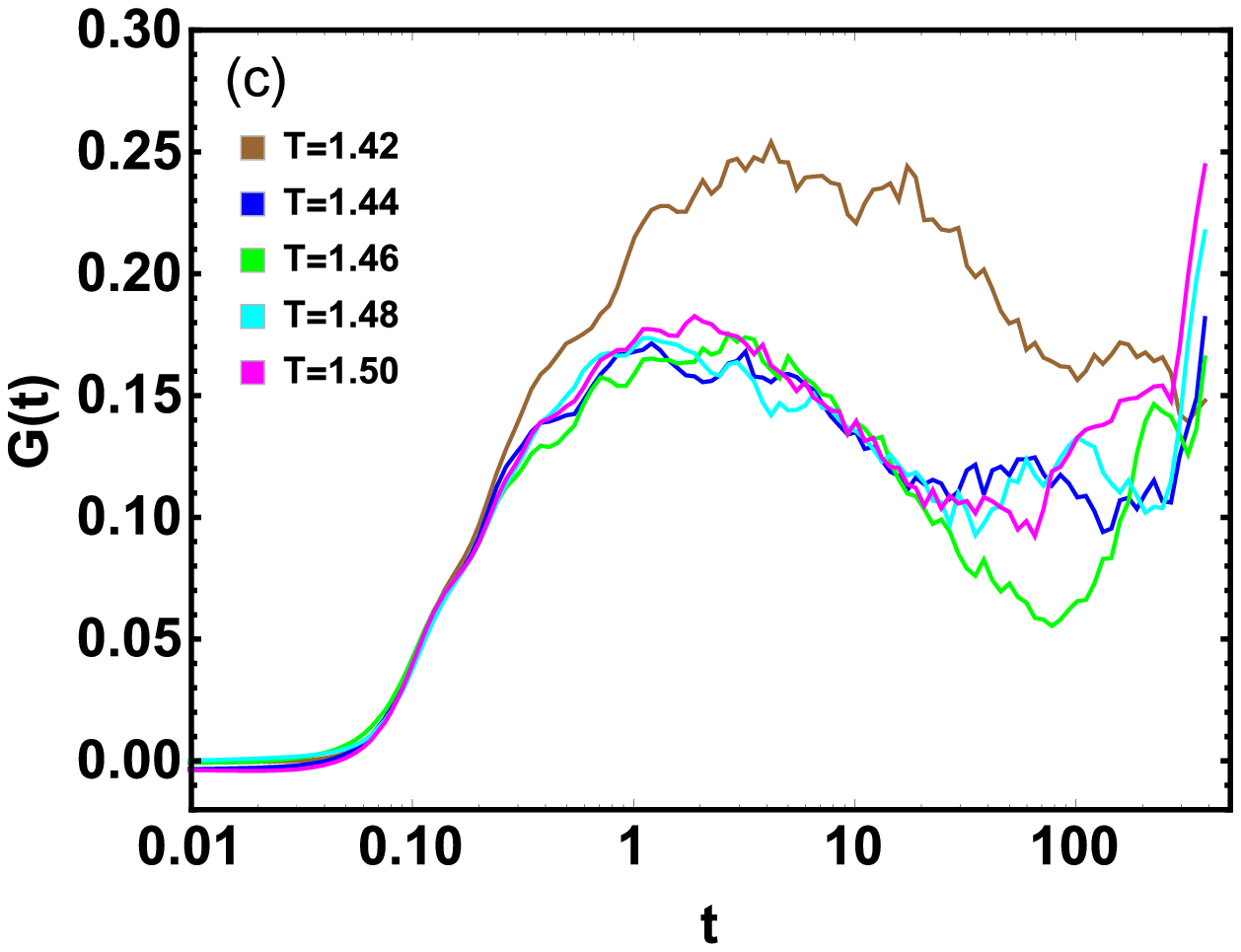}
\caption{Non-Gaussian parameter $G(t)$.  Panel (a): flexible chains.  Panel (b): intermediate-stiffness chains.  Panel (c): stiff chains.  The second peaks at high $t$ for stiff chains are associated with slow rearrangements of the nematic domains. }
\label{fig:ngp}
\end{figure}

The stretched-exponential behavior illustrated in Fig. \ref{fig:sqt}(b) suggests that the dynamics of the $k_b = 4\epsilon$ system are heterogeneous.
We further investigate the potentially differing heterogeneity of dynamics for the different chain stiffnesses by examining the non-Gaussian parameter
\begin{equation}
G(t) = \displaystyle\frac{3\left< r^4(t) \right>}{5 \left< r^2(t) \right>^2} -1
\label{eq:ngp}
\end{equation}
obtained from measurements of diffusion in the melts.
Results for all systems are shown in Figure \ref{fig:ngp}.
For all systems, both the height and the time of the peak in $G(t)$ increase with decreasing $T$.
As expected,\cite{aichele03} the time $\tau_G$ at which $G$ obtains its maximum value $G_{max} \equiv G(\tau_G)$ is comparable to the $\tau_{\alpha}$ obtained from $S(q_{peak},t)$; this corresponds to a crossover from subdiffusive to diffusive behavior at $t \simeq \tau_G$.\cite{donati99}
For flexible and stiff chains, $\tau_G$ and $G_{max}$ remain small even for $T$ near $T_s$, as expected for systems with relatively homogenous dynamics.
Intermediate-stiffness chains show much larger peak values $G_{max}$, much larger $\tau_G$, and a much stronger dependence of $\tau_G$ on $T$, as expected for systems with heterogeneous dynamics.

It is very interesting that the degree of dynamical heterogeneity depends so strongly and nonmonotonically on chain stiffness.
The motion of monomers in intermediate-stiffness chains may be more heterogeneous than that of flexible chains because the angular term in the potential energy favors more cooperative motion.
On the other hand, the more homogeneous motion for stiff chains occurs because they quasi-rigid-rod-like (i.e.\ they move more like rigid rods than their more flexible counterparts.)
Heterogenous dynamics will be further explored in Section \ref{subsec:strings}.

One interpretation of the reason for vitrification in $k_b = 4\epsilon$ systems is that $\tau_{\alpha}$ and $\tau_G$ exceed the time $\tau_0$ over which $T$ crosses the vicinity of $T_s$ during cooling runs.
If one assumes solidification occurs over a range of temperature $\Delta T \simeq .01$ (as is apparent from the widths of the crystallization transitions for $k_b = 0$ and $k_b = 12.5\epsilon$ chains), then $\tau_0 \simeq 10^{-2}/|\dot{T}| = 10^{4} \tau_{LJ}$.
When $\tau_{\alpha} > \tau_0$, systems cannot execute the local rearrangements necessary to convert a liquid to a crystalline state as $T$ decreases from $T_s + \Delta T/2$ to $T_s - \Delta T/2$, and liquid-like disorder gets frozen in as cooling proceeds.
Simulations of cooling at higher $|\dot{T}|$ are consistent with this idea, e.g. $|\dot{T}| = 10^{-4}$ produces vitrification even for $k_b = 0$.

However, this interpretation is not really an explanation, and it is desirable to find a more quantitative and/or microscopic explanation of why $k_b = 4\epsilon$ systems display dynamical slowing-down, dynamical heterogeneity, and glass-formation.
Ref.\ \cite{nguyen15} advanced the simple hypothesis that intermediate-stiffness chains are too flexible to form rodlike configurations but too stiff to form the other bond angles ($\theta = 60^{\circ}, 120^{\circ}$) found in polymeric paths through close-packed crystals with high probability at $T$ near $T_s$, and this ``frustration'' against formation of compatible angles for either RWCP or NCP packing impedes crystallization.

However, this does not explain the non-Arrhenius dynamical slowing down within the liquid state.
We thus turn to a search for other structural signatures that explain it.
Refs.\ \cite{stukalin09,kumar13} predicted that packing frustration increases with chain stiffness, thus increasing structural and dynamical heterogeneity for stiffer systems, which in turn increases fragility (i.e.\ leads to more strongly non-Arrhenius dynamical slowdown.)
One possibility is that this effect is in force for our systems for intermediate $k_b$, but reverses for $k_b$ that are sufficiently large to form nematic melts.
We tested this idea by comparing the dispersion of monomeric Voronoi volumes, $\Delta V_{voro}(T)/\left< V_{voro}(T) \right>$, where $\left< V_{voro} \right>$ and $\Delta V_{voro}$ are respectively the mean and standard deviation of the Voronoi volume distributions, for different $k_b$.
Results for flexible and $k_b = 4\epsilon$ systems were nearly identical for $T > T_s$, indicating that intermediate-chain-stiffness liquids do not have more frustrated packing compared than their flexible-chain counterparts.
This differs from the result of Ref.\ \cite{kumar13}, presumably because that study employed a different angular potential that is minimized at $\theta = 60^{\circ}$ (i.e. employed polymer chains with a ``zigzag'' structure that is more likely to produce packing frustration.)

Another potential source of packing frustration is heterogeneous \textit{cluster-level} structure.\cite{shintani06,malins13a,malins13b}  
In the next subsection we examine this possibility using TCC analysis.

\subsection{TCC analyses}

\begin{figure}[htbp]
\includegraphics[width=2.75in]{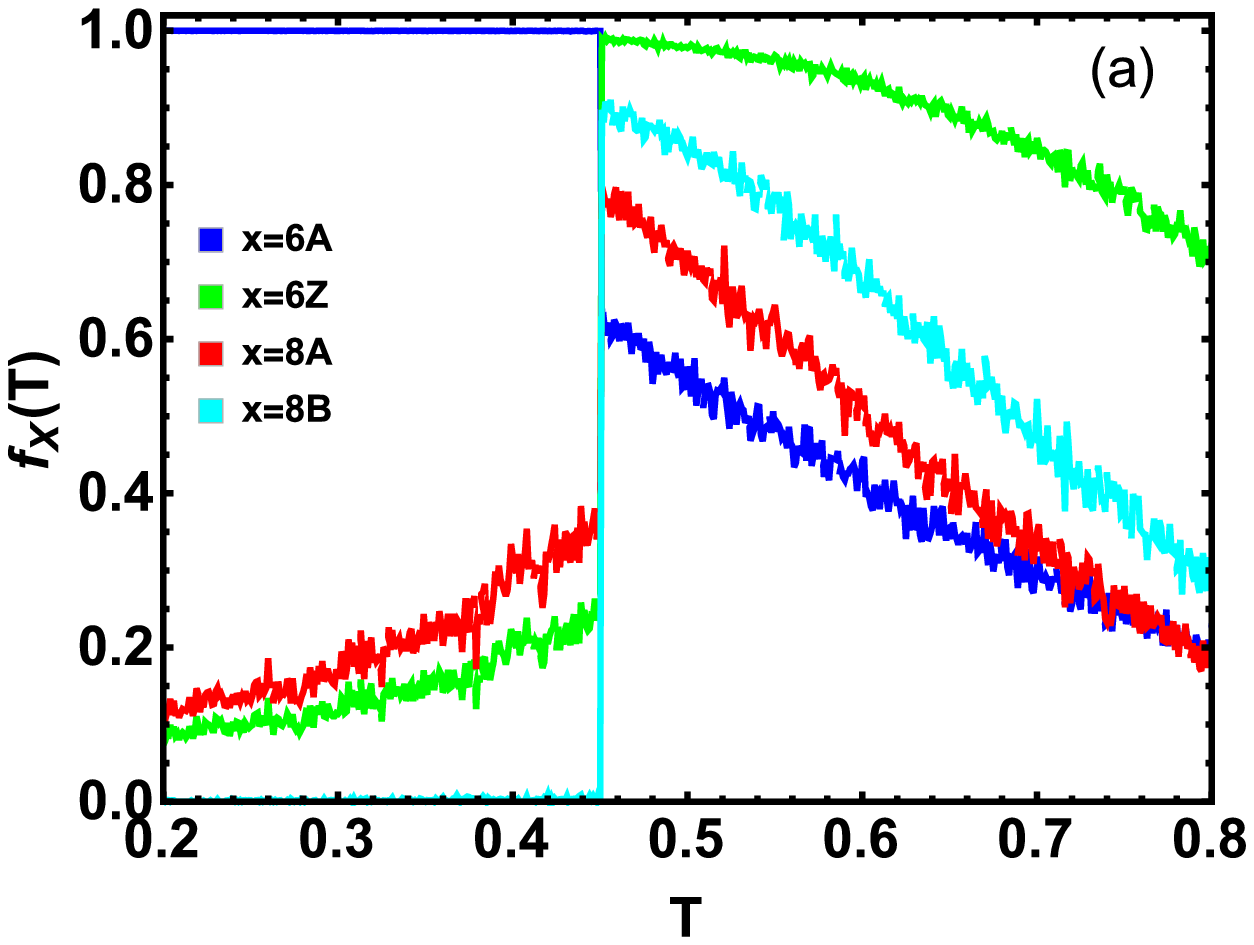}
\includegraphics[width=2.75in]{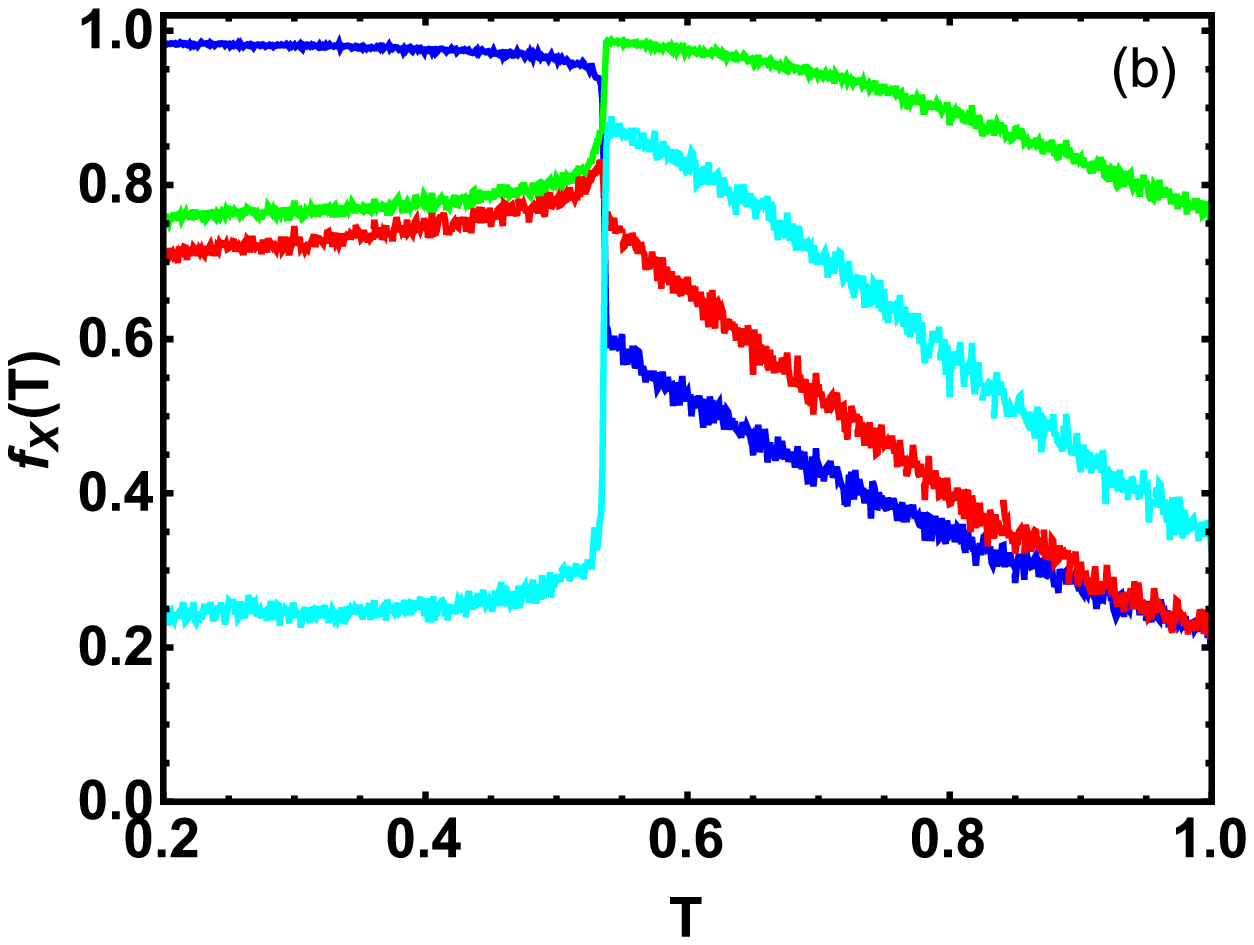}
\includegraphics[width=2.75in]{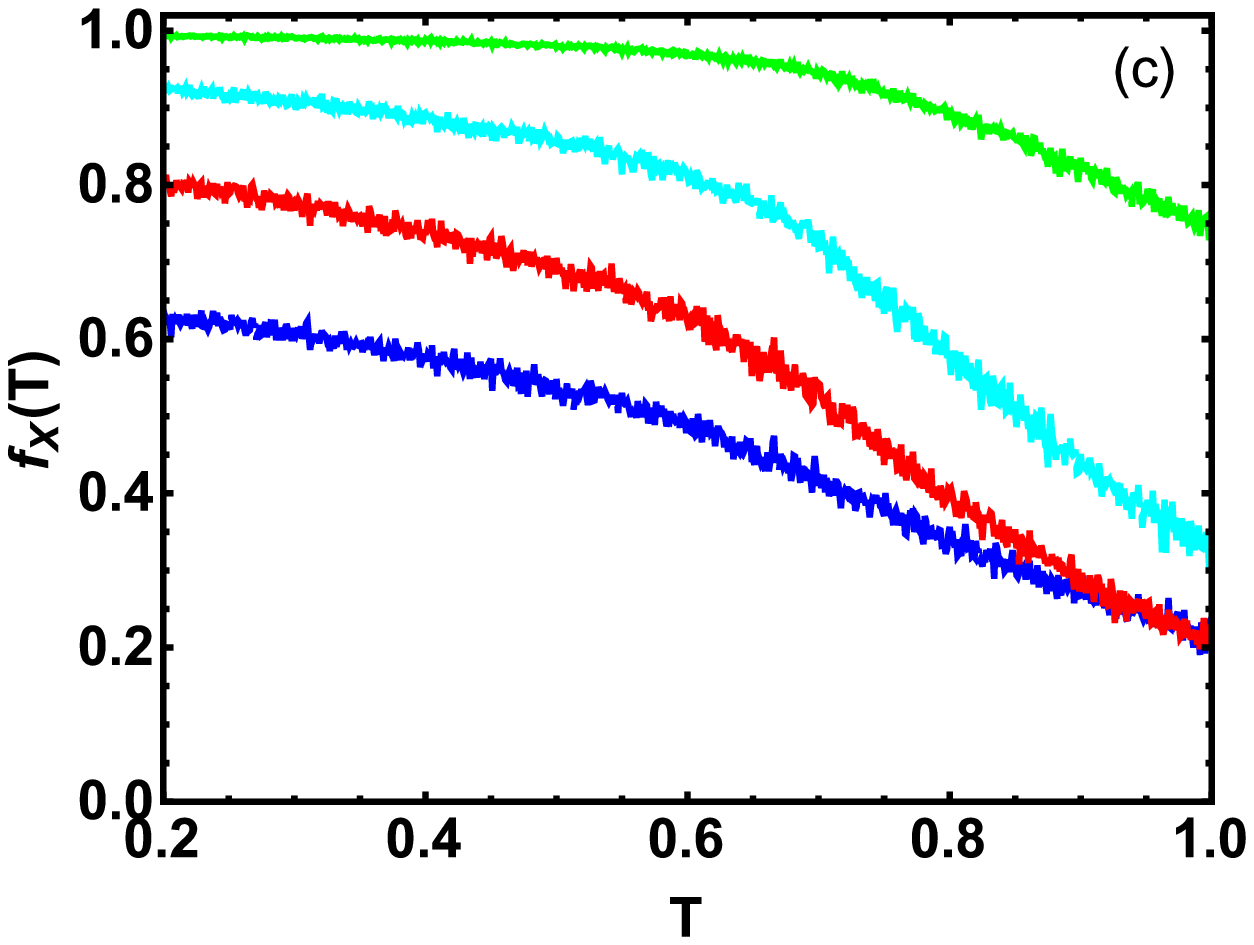}
\includegraphics[width=2.75in]{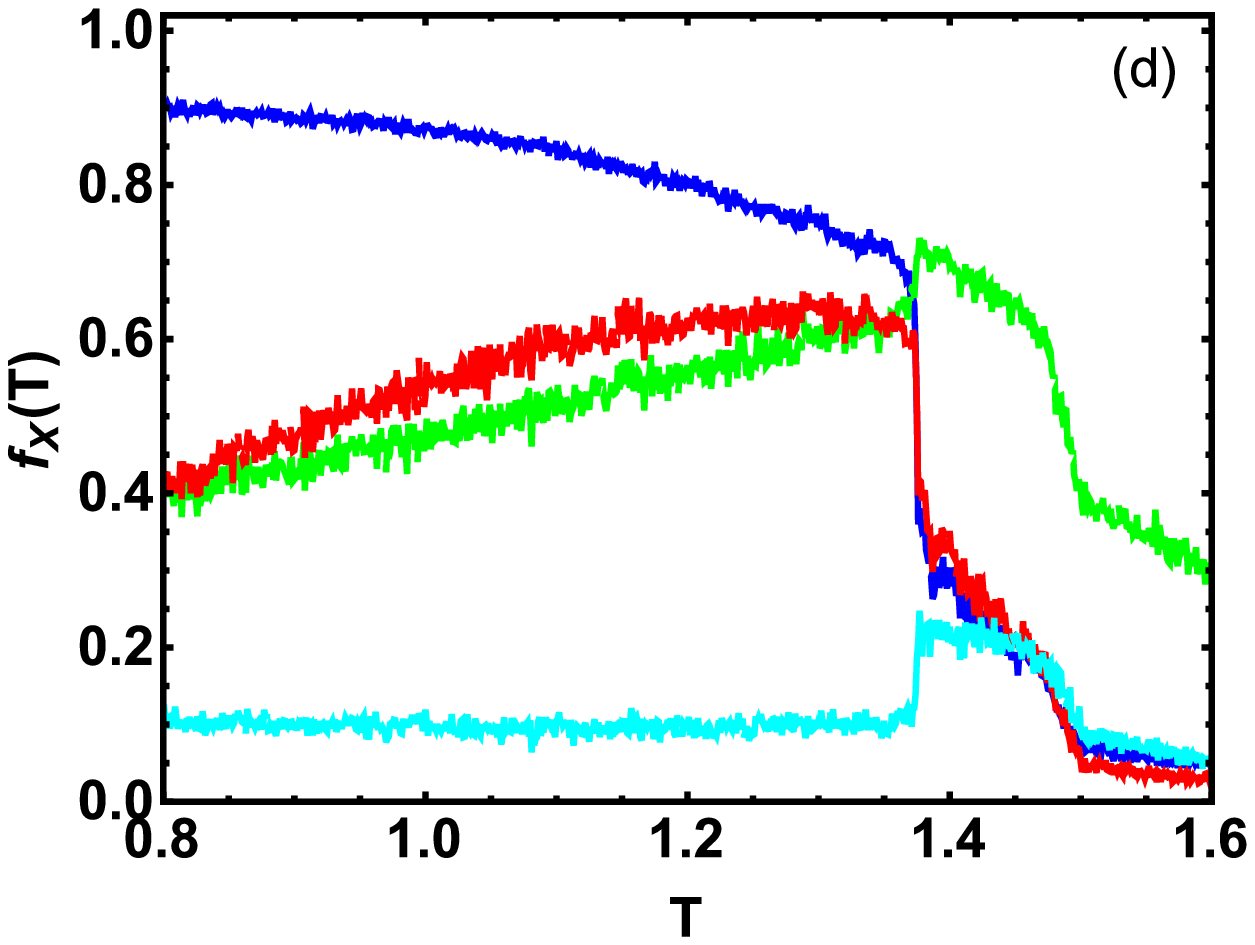}
\caption{Population fractions of particles $f_X(T)$ belonging to the four primary clusters during $|\dot{T}|=10^{-6}/\tau$ quenches.  Panel (a) monomers, (b): flexible chains, (c) intermediate-stiffness chains, (d) stiff chains.  Blue, green, red, and cyan curves respectively indicate data for clusters $x=6A$, $x=6Z$, $x=8A$, and $x=8B$.  Note that the values $f_{X}(T)$ sum to values greater than unity because any particle can be a part of multiple clusters (made up by different sets of other particles.)}
\label{fig:tccpops}
\end{figure}

Figure \ref{fig:tccpops} shows the population fractions of monomers belonging to at least one $6A$, $6Z$, $8A$, or $8B$ cluster, as a function of $T$, for the three chain stiffnesses considered here.  
In addition, panel (a) shows data for monomers.
Note that the monomeric Lennard-Jones system is an excellent crystal-former\cite{trudu06,wang07} which rapidly crystallizes into an FCC structure (with only a few defects) at its $T_s$.
Comparing panel (a) and panel (b), which shows data for flexible chains, shows that the main effect of chains' topological connectivity (in the absence of angular interactions) is raising $T_s$; monomers and flexible-polymer melts show nearly identical values of $f_X$ at the same $T-T_s$.

For all systems, all population fractions $f_X$ increase as $T$ is decreased towards $T_s$; this is because systems' densities (i.e.\ $\phi$) are increasing, whereas the cutoff radius $r_{TCC}$ used to identify neighboring monomers in the TCC analyses\cite{malins13d} is $T$-independent.
Note that all systems show a significant degree of locally octahedral ordering $f_{6A}$ even well above $T_s$; the presence of these subcritical nuclei is typical for crystallizable systems that develop MRCO well above solidification.\cite{shintani06,russo12}
Upon solidification, $f_X$ for amorphous clusters ($6Z$, $8A$, and $8B$) drops sharply for systems that crystallize.
For these systems, locally crystalline order as measured by $f_{6A}$ increases sharply at the same time.
The drop in amorphous-cluster population fractions is less pronounced for flexible polymers than for monomers because chain connectivity restricts polymers' ability to locally rearrange, e.g.\ rearrangements of $6$-atom clusters from $6Z$ into $6A$ order are hindered by chain backbone uncrossability.

In general, values of $f_X$ in flexible and stiff-chain systems exhibit first-order-like transitions upon crystallization.
This is expected; what is surprising is that intermediate-stiffness chains behave so differently.
For $k_b = 4\epsilon$ systems, all $f_X$ continue increasing steadily as $T$ is decreased past $T_s$, with no significant change in their slopes.
This is so despite the fact that $k_b = 4\epsilon$ systems, like their counterparts for all $k_b > 0$, possess a crystalline (NCP) ground state.
Perhaps more surprisingly, it is so despite the fact that flexible and intermediate-stiffness systems are similarly structured at the level of single clusters, i.e.\ their values of $f_X$ and ratios $f_X/f_Y$ are similar at $T$ slightly above $T_s$.

Panel (d) of Figure \ref{fig:tccpops} shows that stiff-chain melts possess rather different cluster-level structure than their flexible-chain counterparts.
The nematic ordering of the melts suppresses fivefold-symmetric order; instead, hexagonal-like order exists in the planes perpendicular to the nematic director field.
This may be part of the reason why these systems are good crystal-formers.
Nonetheless, the main conclusion from this panel is that differences in cluster-level structure between stiff and more-flexible melts are greater than the corresponding differences between intermediate-stiffness and flexible melts (panels (b-c)).
Thus it is difficult to attribute the dynamical arrest to differences in the cluster population fractions $f_X$ or their ratios.

For all $k_b$ for temperatures above solidification, there are large populations of mutually incommensurable clusters (e.g.\ $6A$ and $6Z$).  
This suggests a large kinetic bottleneck for crystallization; the incommensurability must be alleviated for the melts to crystallize.  
For flexible and stiff chains it is alleviated during solidification (i.e. $f_{6Z}$ drops sharply at $T_s$), while for intermediate-stiffness chains this does not happen.  
In other words, significant packing frustration exists in the melt state for all $k_b$, and is alleviated upon solidification for flexible and stiff chains but not for intermediate-stiffness chains.  
The question again raised is: why is this so?

Refs.\ \cite{malins13a,malins13b} showed that the dynamical slowdown in model colloidal glass-formers is associated with percolation of the amorphous clusters; mean lifetimes of these clusters increase sharply with decreasing $T$ as their populations increase.
One might expect this to also be true in our systems, but it does not seem to be; examination of snapshots of various amorphous-ordered clusters shows no obvious difference in amorphous-cluster percolation levels between flexible and intermediate-stiffness systems at similar values of $T-T_s$.\cite{footperc}

\begin{figure}[htbp]
\includegraphics[width=2.75in]{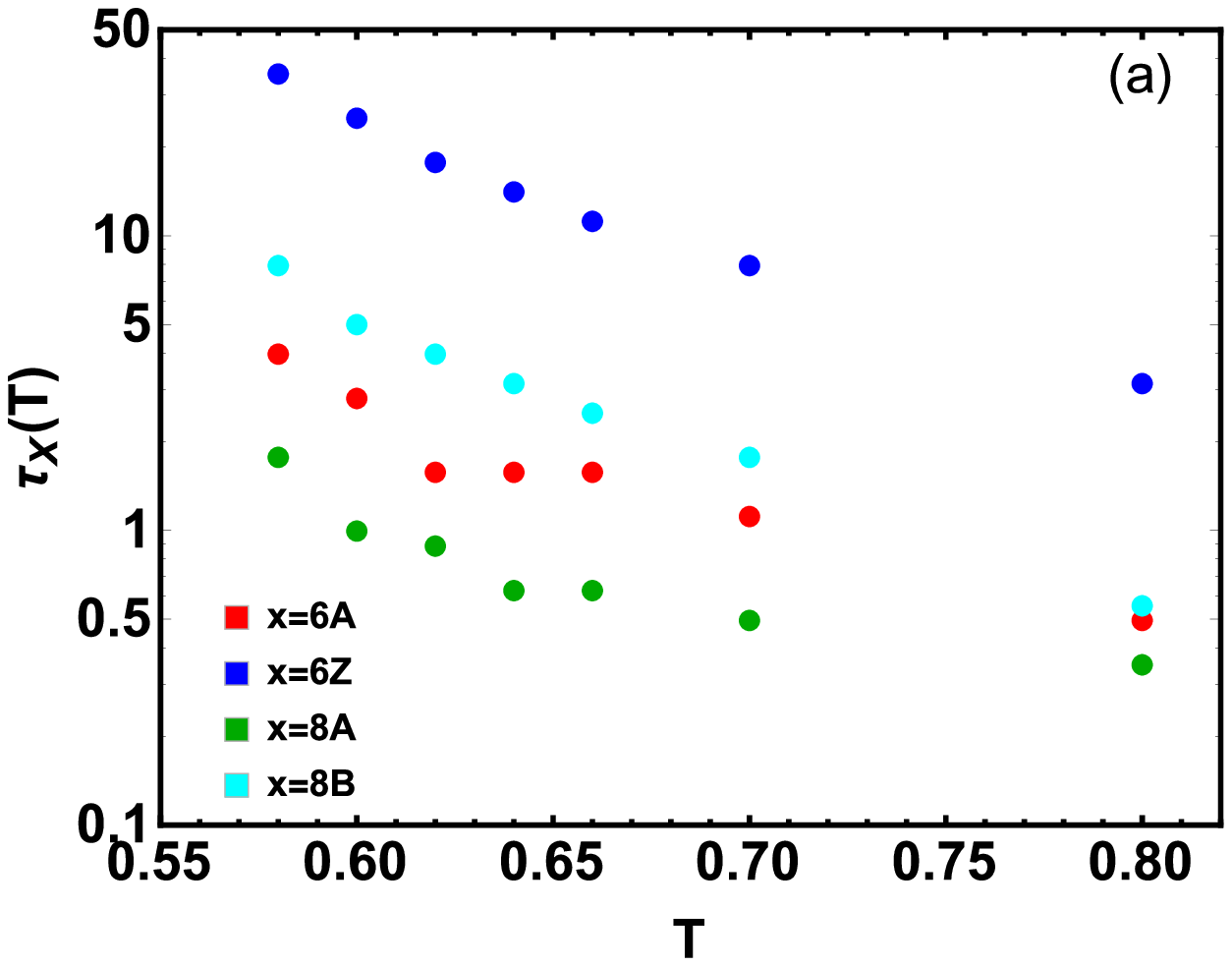}
\includegraphics[width=2.75in]{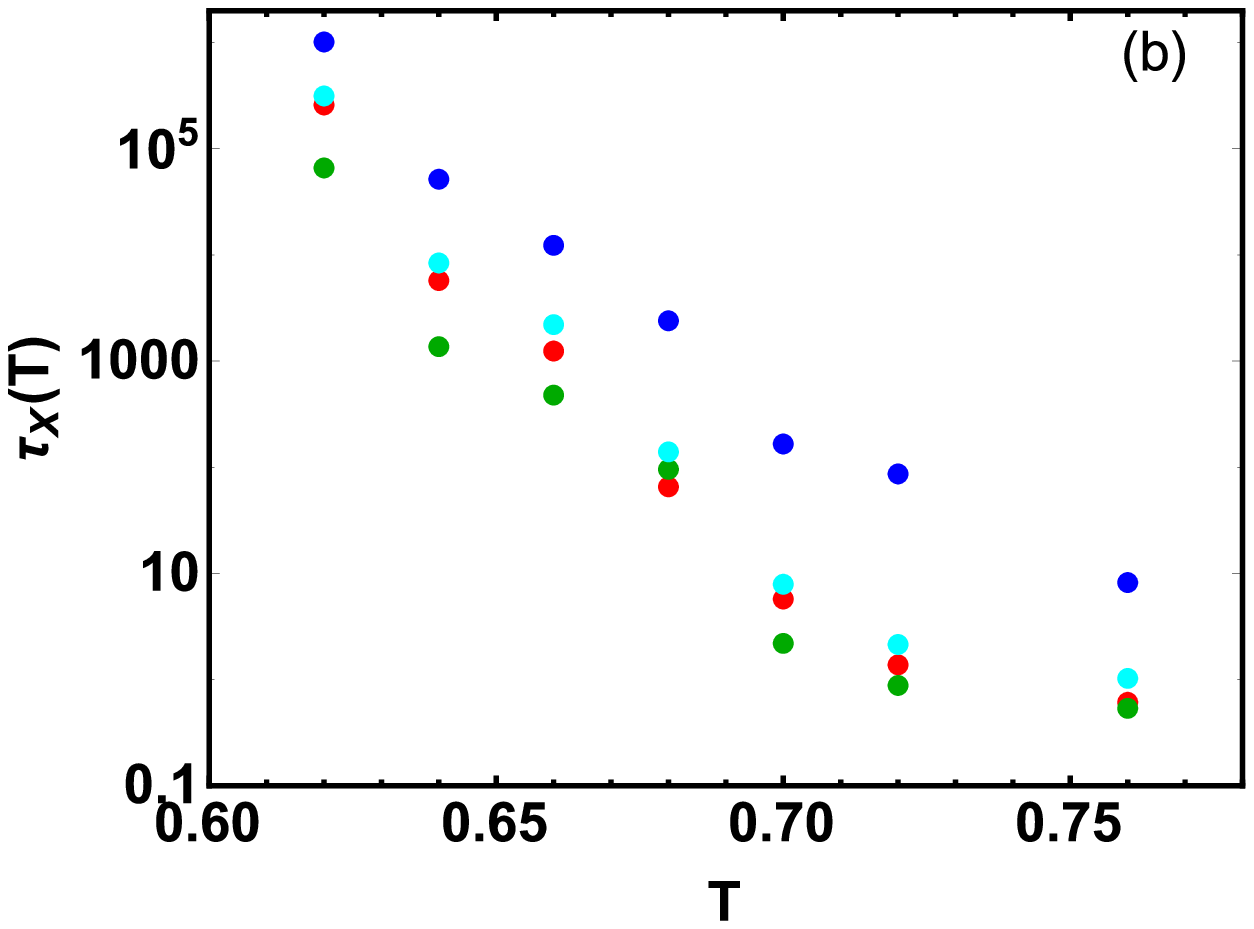}
\caption{Values of $\tau_{6A}$, $\tau_{6Z}$, $\tau_{8A}$ and $\tau_{8B}$ versus temperature, in systems of flexible (panel a) and intermediate-stiffness (panel b) chains.}
\label{fig:clustlife}
\end{figure}

Another potential answer is that  the abovementioned dynamical heterogeneity is closely associated with this heterogeneous cluster structure.
Transient structural ordering has been extensively linked to dynamical heterogeneity in recent years.\cite{kawasaki07,tanaka10,royall15}
Recent simulation studies have found that regions of locally icosohedral\cite{malins13a,malins13b} (and in other systems, crystalline\cite{shintani06,leocmach12}) order are dynamically slower than their more ordered (more amorphous) counterparts.
Figure \ref{fig:clustlife} indicates the lifetimes $\tau_X$ for the four clusters of primary interest, calculated by identifying $A_X(2\tau_X) = 1/e^2$ in Eq.\ \ref{eq:aoft}.\cite{why2tau}
The 8-particle clusters naturally have shorter lifetimes than their 6-particle counterparts because for our definition of $A_X(t)$ (Eq.\ \ref{eq:aoft}) there are more ways for larger-$N$ clusters to decay, i.e.\ by any of the $N$ particles in the cluster hopping away. 
Data for stiff chains are not shown because the values of $\tau_X$ are very small ($\lesssim 10\tau_{LJ}$) and their trend with $T$ is not clear at the high temperatures considered.

For flexible and intermediate-stiffness chains, comparing data for $\tau_{6A}$ to data for $\tau_{6Z}$ and data for $\tau_{8A}$ to data for $\tau_{8B}$ provides a partial explanation of the heterogeneous dynamics.
Clusters with fivefold or partial-icosohedral order are more stable in the liquid state and have larger lifetimes, as expected.
The associated slower structural of relaxation of regions with more liquidlike ordering helps explain the stretched-exponential relaxation observed for $S(q_{peak},t)$.
Values of $\tau_{6Z}$ are comparable to values of $\tau_{\alpha}$ and exhibit Vogel-Fulcher-like temperature dependence for $k_b = 4\epsilon$ chains, whereas for flexible chains they show a more Arrhenius $T$-dependence.

The larger lifetimes of clusters that are subsets of icosohedra (i.e.\ $\tau_{6Z} > \tau_{6A}$ and $\tau_{8B} > \tau_{8A}$) are consistent with previous results\cite{malins13a,malins13b} indicating such clusters play a key role in glass-formation for some systems.
However, $\tau_{6Z} > \tau_{6A}$ also holds for flexible-chain systems that possess ``fast, simple'' dynamics.
Furthermore, while one might expect the ratio $\tau_{6Z}(T)/\tau_{6A}(T)$ to increase sharply as $T_s$ is approached in a glassforming system, in fact it depends only weakly on temperature.
The similar behavior in Fig.\ \ref{fig:clustlife} for glass-forming and crystallizing systems may cast some doubt on the generality of the conclusions reached by studies (such as Refs.\ \cite{malins13a,malins13b}) that analyzed glass-formation in terms of the differences in cluster lifetimes in systems interacting via a single potential.
Future studies of the CF-GF competition may be enhanced by comparing results for different interaction potentials, as was done in Refs.\ \cite{shintani06,royall15}.

\subsection{Other measures of structure}

Examination of other structural metrics reinforces the conclusion that the dynamical arrest reported above for intermediate-stiffness chains is quite difficult to relate to any static-structural signature.
The two-body excess entropy (i.e.\ the difference between the entropy of a system and that of an ideal gas at the same temperature and density),
\begin{equation}
s_2 = -\displaystyle\frac{k_B \rho}{2} \int 4\pi r^2 \left[g(r) \ln{g(r)} - (g(r)-1) \right] dr,
\label{eq:excess}
\end{equation}
where $g(r)$ is the pair correlation function, was shown in Ref.\ \cite{mittal06} to correlate with the temperature and density dependence of diffusivity in model glass-forming liquids.
Lower values of $s_2$ were associated with both more-ordered liquid structure and lower diffusivity (i.e. larger relaxation times.)
One might therefore expect intermediate-stiffness liquids to possess lower values of $s_2$ at similar $T-T_s$.
However, examination of $s_2(T)$ for our systems shows that results for flexible and intermediate-stiffness chains are essentially identical for $T > T_s + .05$.
$s_2$ is very slightly larger (higher: $-s_2$ is smaller) for $k_b=4\epsilon$ for $T$ very near $T_s$, consistent with $k_b = 0$ liquids being slightly more ordered.
However, this difference is small compared to the difference with stiff-chain systems (which possess a significantly larger $-s_2$ arising from their additional, nematic order), and is therefore difficult to associate with the dynamical arrest.

Alternatively, one might imagine that nematic order in the intermediate-stiffness system is more heterogeneous, and that the presence of regions of higher and lower $\mathcal{O}$ produces frustration leading to the dynamical slowdown.
However, this is not the case; the dispersion $\Delta \mathcal{O}(T)/\left< \mathcal{O}(T) \right>$ is nearly identical for flexible and intermediate-stiffness systems above $T_s$ (similarly to the abovementioned dispersion of Voronoi volumes.)

Finally, multiple studies have related crystallizability to the propensity for development of orientational order.\cite{shintani06,kawasaki07,tanaka10,leocmach12,russo12} 
The same studies have linked regions of high orientational order (MRCO) to locally slow dynamics.
Following these works, we compared results for the Steinhardt order parameter\cite{steinhardt83} $Q_6(T)$ for our systems.
Results were similar to those for the Voronoi-volume and $\mathcal{O}$ distributions: $\left< Q_6(T) \right>$ is nearly identical for flexible and intermediate-stiffness-chains for $T > T_s$, eliminating different bond-orientational order as the cause of the dynamical arrest.

\subsection{Stringlike cooperative motion}
\label{subsec:strings}

\begin{figure}[htbp]
\center
\includegraphics[width=2.25in]{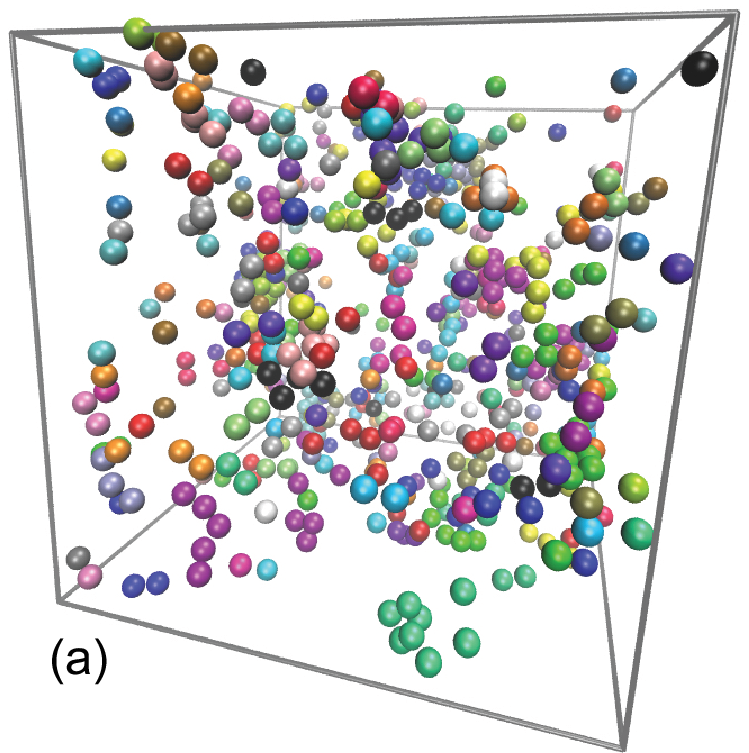}
\includegraphics[width=2.25in]{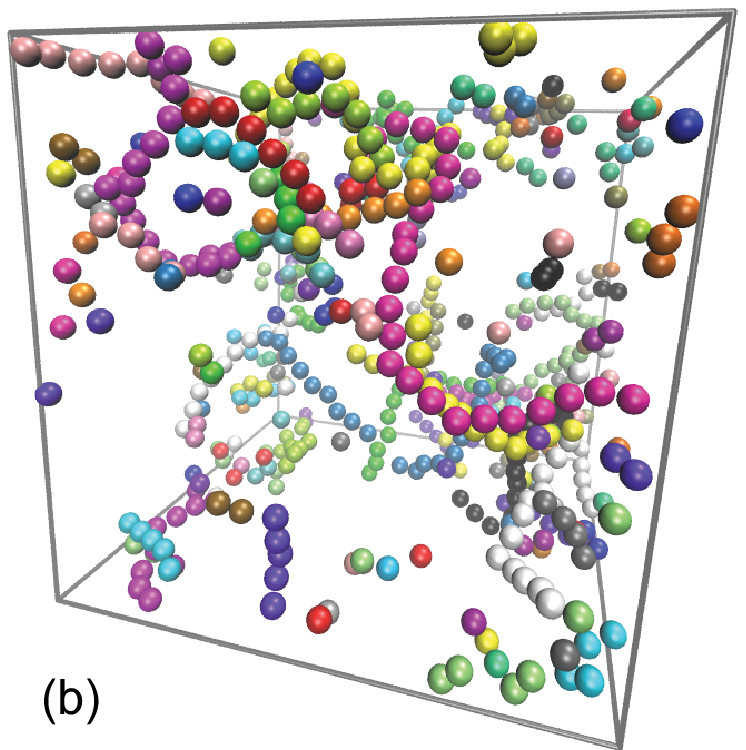}
\includegraphics[width=2.75in]{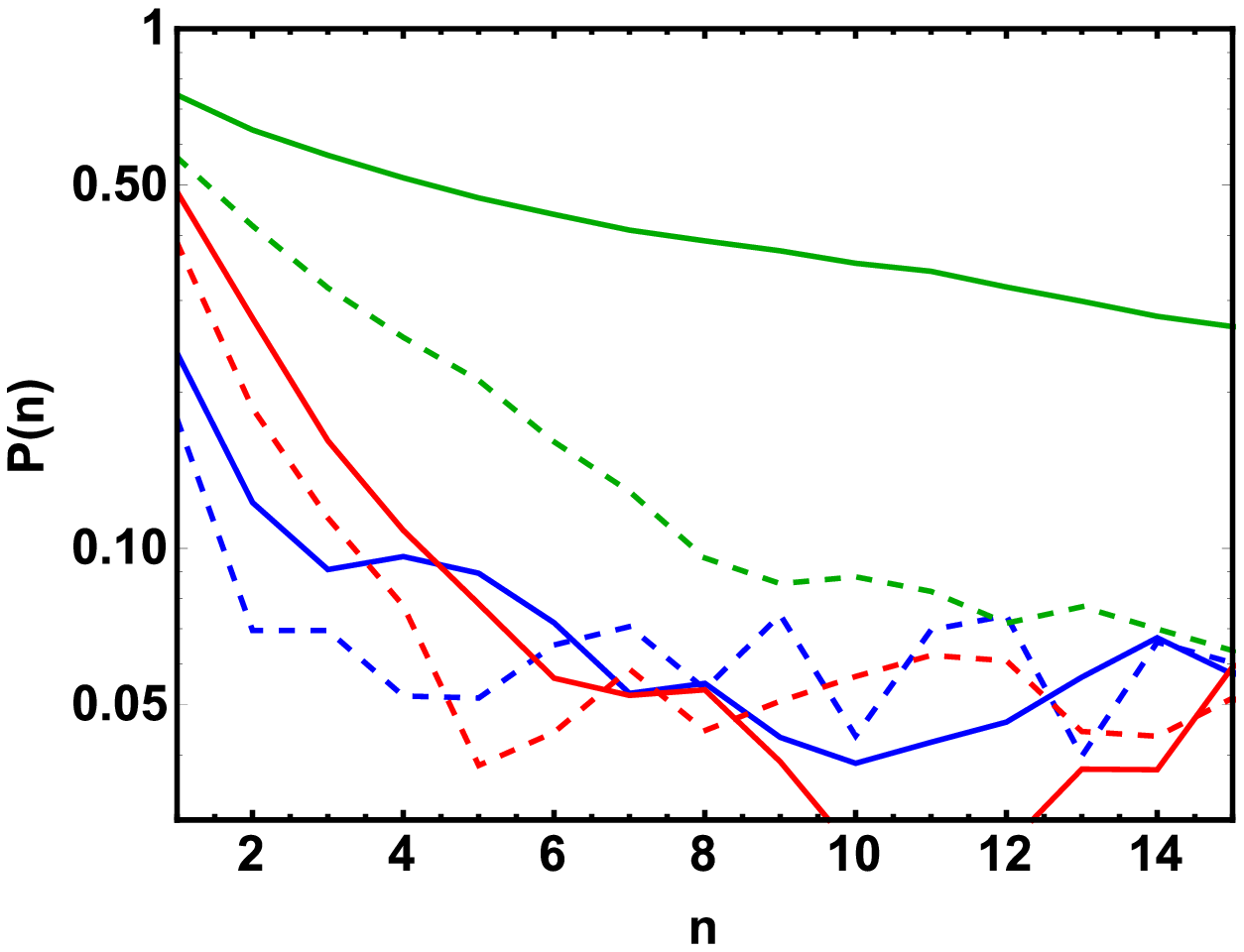}
\caption{Stringlike particle motion.  Panels (a-b) show strings for the time intervals $t' \in (0, \tau_G)$ at $T$ slightly above $T_s$, respectively $T = 0.58$ for $k_b = 0$ (panel a) and and $T=0.62$ for $k_b = 4\epsilon$ (panel b).  Images were generated using VMD.\cite{vmd01}  Panel (c) shows $P(n)$ for $k_b = 0$, (blue curves) and $k_b = 4\epsilon$ (green curves) for the abovementioned $T$ slightly above $T_s$ (solid curves) and for $T=0.80$ (dashed curves).  Data for $k_b = 12.5\epsilon$ are shown in red, for $T=1.42$ (solid curve) and $1.50$ (dashed curve).}
\label{fig:strings}
\end{figure}

Another way to interrogate dynamical slowdown is to examine spatial correlations of particle mobility.
Many studies\cite{donati98,donati99,aichele03,starr13} have shown that the sets of particles which are most mobile over timescales corresponding to maximally heterogeneous dynamics form noncompact, ``stringlike'' structures.
Roughly speaking, the strings are formed when one particle undergoes a hop-like displacement, another particle hops into the space it leaves behind, and so on.
These studies have shown that  the length of the strings increases as $T_g$ is approached from above, and it is now generally agreed upon\cite{starr13} that the strings and stringlike motion are examples of the ``cooperatively rearranging regions'' of Adam and Gibbs,\cite{adam65} and thus play a key role in controlling the glass transition.

We now examine how chain stiffness affects potentially stringlike motion in our systems.
We follow the procedure used in Ref.\ \cite{donati99} to identify ``highly mobile'' particles as the $5\%$ of particles undergoing the largest displacements over a time interval $\tau_G(T)$.
In other words, highly mobile particles are those which move the furthest over the time interval over which dynamics are maximally heterogeneous.
Figure \ref{fig:strings} shows snapshots of these particles for temperatures just above solidification: $T=0.58$ for $k_b = 0$ (panel a) and $T=0.62$ for $k_b = 4\epsilon$ (panel b).
The correlations of mobile particles are much more obviously stringlike for the glassforming, intermediate-stiffness system.

Different colors in Figure \ref{fig:strings}(a-b) indicate monomers belonging to different chains.
One can clearly see that the mobile-particle sets (hereafter referred to as strings) for intermediate-stiffness systems correspond much better to chain backbones than the strings for flexible-chain systems; nearby mobile particles for the latter are far more likely to be spread amongst multiple chains.
In other words, for intermediate-stiffness chains (but much less so for flexible and stiff chains) the strings often correspond to large sections of chains executing coordinated motion along their backbones.
Panel (c) illustrates this quantitatively by plotting $P(n,T)$, the probability that monomers a chemical distance $n$ away from a mobile monomer on the same chain are also mobile at temperature $T$. 
Random mobility of monomers along chains would produce $P(n,T)=0.05$.
Actual mobility correlations are short-ranged for flexible and stiff chains, but long-ranged for intermediate-stiffness chains.
The correlations of particle mobility along chains also increase with decreasing $T$, and do so more strongly for $k_b = 4\epsilon$ chains.

We believe that this effect is the source of dynamical arrest in the intermediate-stiffness systems.
Monomer hops can more easily occur in directions perpendicular to the chain backbone for flexible chains than for intermediate-stiffness chains, because the angular energy term $U_{bend}(\theta)$ (Eq.\ \ref{eq:Ubend}) imposes an energy cost for such hops.
On the other hand, few excursions away from $\theta_i=0$ occur for stiff-chain systems, and mobility distributions are narrower since these chains undergo quasi-rigid-rod-like motion.
The net effect is that monomer hops for intermediate-stiffness chains (but \textit{not} flexible or stiff chains) are apparently an activated process that induces cooperative motion; when one monomer hops, it pulls its (chemically) nearby intrachain neighbors along with it.
This motion resembles ``crawling.''
The Vogel-Fulcher relaxation observed for these systems could potentially arise from an increase in the activation energy for hops as $\rho$ increases, together with the increase in hop correlation along chain backbones.

\section{Discussion and Conclusions}

In this paper, we have analyzed the coupled chain-stiffness and temperature dependence of both dynamics and microstructure in model crystallizable bead-spring polymer melts.
We found nonmonotonic dependence of dynamics upon chain stiffness; both flexible and stiff chains possess ``fast, simple'' (Arrhenius) dynamics, whereas intermediate-stiffness chains exhibit the dynamics of fragile glass-formers.
This result complements previous simulation studies (e.g.\ Refs.\ \cite{faller99,faller01,kumar13}) that examined the dependence of melt dynamics on chain stiffness.
For example, Ref.\ \cite{kumar13} found that fragility (i.e. dynamical slowdown) increases with chain stiffness, while here we showed that this effect is nonmonotonic and reverses when melts become nematically ordered.

Our attempts to isolate a microscopic static-structural cause of the different dynamics yielded no clear ``smoking gun.''
Indeed, predicting whether a system will be a glassformer in terms of its interactions and microstructure is well-known as an extremely difficult problem.\cite{debenedetti01,malins13a,royall15}
However, the different gross dynamics \textit{are} clearly linked to qualitatively different heterogenous monomer-scale dynamics.
For intermediate-stiffness chains, stringlike motion\cite{donati98} corresponds to activated ``crawling'' along chain backbones.
Such crawling is far less prominent for both flexible and stiff chains.
Previous studies\cite{aichele03} that found mobile-particle strings to be largely uncorrelated with chain backbones employed fully flexible chains; our result suggests an additional mechanism of activated/cooperative rearrangement for intermediate-stiffness chains.

Many interesting simulation studies of polymer crystallization have appeared recently.\cite{luo09,sommer10,luo13,luo14,yi09,yi13,anwar13}
Nearly all of these have employed atomistic or united-atom models to study specific polymer chemistries.
Such studies certainly can identify phenomena which are general to many different polymers, but their use of single interaction potentials rather than comparing behavior for a range of potentials often obscures this generality.  
Furthermore, few of these studies have connected solidification behavior directly to temperature-dependent steady-state melt dynamics, and none have connected it to the chain stiffness dependence of these dynamics.
Here we have done so for unentangled chains.
Extension of this work to entangled systems would be very challenging since the stiffness dependence of the disentanglement dynamics\cite{faller01,luo13,luo14} will couple to the CF-vs.-GF-related dynamics described above, but would be a worthy goal.

Experimentally observing local microstructural motifs comparable to the clusters discussed herein may not be possible for typical polymers, due to the small length scales and short time scales involved.
However, such motifs have been observed in colloidal systems\cite{gasser01} using confocal microscopy, and recent studies have also examined their relaxation dynamics.\cite{leocmach12,taffs13}
Variable-stiffness colloidal and granular polymers\cite{zou09,feng13} have recently been synthesized, and it would be interesting to study the CF-GF competition in these systems.

This material is based upon work supported by the National Science Foundation under grant no.\ DMR-1555242.
We gratefully acknowledge Monojoy Goswami for helpful discussions.


\begin{thebibliography}{59}
\bibitem[{col()}]{collpol}
\bibinfo{note}{Ref.\ \cite{taffs10b} applied TCC to model colloid-polymer
  mixtures, but employed simple Asakura-Ooosawa interactions wherein polymers
  are treated as ideal Gaussian coils lacking excluded volume interactions or
  chain stiffness.}
\bibitem[{why()}]{why2tau}
\bibinfo{note}{The factor-of-2 is chosen because $A_X(t)$ displays an extreme
  2-step behavior wherein a fraction comparable to $1-1/e$ of clusters decay
  relatively rapidly.\cite{malins13b} Because this fast step produces decay of
  $A$ to values in the range $0.3-0.4$ (i.e.\ both below and above $1/e$,
  depending on which system is considered), the definition of $\tau_X$
  including the factor of 2 produces better behaved results for $\tau_X$ than
  does $A_X(\tau_X) = 1/e$.}
\bibitem[{foo({\natexlab{a}})}]{footTs}
\bibinfo{note}{Values of $T_s$ were determined using both the transitions in
  Fig.\ \ref{fig:grossvsT} and the $NPT$ simulations. If crystallization
  occurred at a higher $T$ in the latter, that higher value is quoted here for
  $T_s$.}
\bibitem[{foo({\natexlab{b}})}]{footperc}
\bibinfo{note}{Another possibility is that effective percolation is achieved by
  connecting different amorphous-cluster regions through chain backbones. This
  is hard to either verify or rule out.}
\end{thebibliography}

\end{document}